\documentclass[a4paper,12pt]{amsart}
\usepackage[foot]{amsaddr}

\usepackage{amssymb}
\usepackage{siunitx}
\usepackage{amsfonts}
\usepackage{booktabs}
\usepackage{amsmath,array}
\usepackage{caption}
\usepackage{subcaption}  
\usepackage{tabularx}
\usepackage{graphicx}
\usepackage[margin=2.7cm]{geometry}
\usepackage{bm}  
\usepackage{hyperref}

\begin{document}

\title[A ROM of a turbulent buoyant flow over a backward-facing step]{A POD-Galerkin reduced order model of a turbulent convective buoyant flow of sodium over a backward-facing step}

\author{Kelbij Star\textsuperscript{1,2*}}
\address{\textsuperscript{1}SCK$\cdot$CEN, Institute for Advanced Nuclear Systems, Boeretang 200, 2400 Mol, Belgium.}
\address{\textsuperscript{2}Ghent University, Department of Electromechanical, Systems and Metal Engineering, Sint-Pietersnieuwstraat 41, B-9000 Ghent, Belgium.}
\thanks{\textsuperscript{*}Corresponding Author.}
\email{kelbij.star@sckcen.be}

\author{Giovanni Stabile\textsuperscript{3}}
\address{\textsuperscript{3}SISSA, International School for Advanced Studies, Mathematics Area, mathLab, via Bonomea 265, 34136 Trieste, Italy.}

\email{gstabile@sissa.it}

\address{\textsuperscript{4}FlandersMake@UGent.}

\author{Gianluigi Rozza\textsuperscript{3}}
\email{grozza@sissa.it}

\author{Joris Degroote\textsuperscript{2,4}}
\email{Joris.Degroote@UGent.be}


\date{}

\dedicatory{}

\begin{abstract} 
A Finite-Volume based POD-Galerkin reduced order modeling strategy for steady-state Reynolds averaged Navier--Stokes (RANS) simulation is extended for low-Prandtl number flow. The reduced order model is based on a full order model for which the effects of buoyancy on the flow and heat transfer are characterized by varying the Richardson number. The Reynolds stresses are computed with a linear eddy viscosity model. A single gradient diffusion hypothesis, together with a local correlation for the evaluation of the turbulent Prandtl number, is used to model the turbulent heat fluxes. The contribution of the eddy viscosity and turbulent thermal diffusivity fields are considered in the reduced order model with an interpolation based data-driven method. The reduced order model is tested for buoyancy-aided turbulent liquid sodium flow over a vertical backward-facing step with a uniform heat flux applied on the wall downstream of the step. The wall heat flux is incorporated with a Neumann boundary condition in both the full order model and the reduced order model. The velocity and temperature profiles predicted with the reduced order model for the same and new Richardson numbers inside the range of parameter values are in good agreement with the RANS simulations. Also, the local Stanton number and skin friction distribution at the heated wall are qualitatively well captured. Finally, the reduced order simulations, performed on a single core, are about $10^5$ times faster than the RANS simulations that are performed on eight cores.
\end{abstract}

\maketitle
\section{Introduction} 
Heat transfer in liquid metals and molten salts is of interest, for instance, in nuclear facilities that use high-temperature heat transfer media, so-called low-Prandtl (Pr) fluids, as a coolant. Due to the high thermal diffusivity of low-Pr fluids, where Pr is the ratio of diffusivity of momentum to diffusivity of heat, the influence of buoyancy on the flow field is present at much higher Reynolds numbers compared to air or water~\cite{taler2016heat}. Therefore, the flow regime between forced and natural convection, where driven flow interacts with buoyancy effects, needs to be studied in many heat transfer applications~\cite{niemann2017turbulence}. 

Low-Prandtl number fluid turbulent flows, and especially their associated turbulent heat fluxes, are complicated to model numerically as heat conduction through the boundary layer has more dominant effect with respect to convection. Therefore, the thermal boundary layers become thicker when the Prandtl number is decreased. This means that there is a difference in the range of the spatial (and temporal) scales of temperature and velocity. As a consequence, the conductive heat fluxes near walls become more important. Therefore, it is problematic to apply the Reynolds analogy, which assumes a constant turbulent Prandtl number, Pr$_t$, close to unity, to calculate the local turbulent heat fluxes~\cite{roelofs2015status}. Furthermore, Pr influences not only the temperature field and the heat flux modeling, but also the velocity field and the shear modeling in the case of buoyancy-aided flows~\cite{grotzbach2013challenges}. Therefore, heat transfer in liquid metals, compared to common fluids with a Prandtl number around unity, requires additional or different (physical) modeling.

Only a few numerical studies on incompressible turbulent convective buoyant flows for low-Prandtl number fluid flows can be found in literature. Three studies are highlighted here: Cotton and Jackson~\cite{cotton1990vertical} performed numerical calculations for a buoyancy-aided mixed convective turbulent flow in a vertical pipe for liquid sodium (Pr = 0.005-0.01). Niemann and Frohlich~\cite{niemann2017turbulence} investigated a turbulent flow of liquid sodium over a backward-facing step at forced and buoyancy-aided mixed convection using direct numerical simulation (DNS). And most recently, Oder et al. presented direct numerical simulation of low-Prandtl fluid flow over a confined backward-facing step~\cite{oder2019direct}.

Schumm et al.~\cite{schumm2016numerical,schumm2018investigation} compared steady-state Reynolds-averaged Navier--Stokes \newline (RANS) simulations with the direct numerical simulations performed by Niemann and Frohlich~\cite{niemann2017turbulence} and concluded that the predicted velocity, turbulence kinetic energy and Reynolds shear stress profiles are in good agreement with the DNS data. They based the choice of the turbulence model for the Reynolds stresses, namely the Ince and Launder's model~\cite{ince1989computation}, on the study of Cotton and Jackson~\cite{cotton1990vertical}. This turbulence model is basically the model of Launder and Sharma~\cite{launder1974application} including the near-wall length-scale correction term from Yap~\cite{yap1987turbulent} in the equation of the dissipation rate of turbulence kinetic energy. The model is widely used due to its algorithmic simplicity and relatively good performance~\cite{roelofs2015status,hsieh2004numerical} compared to the more advanced model of Hanjali{\'c} et al.~\cite{hanjalic1996natural} and second-moment closure models (e.g. Craft et al.~\cite{craft1996recent}, Dol et al.~\cite{dol1999dns} and Manceau et al.~\cite{manceau2000turbulent2}). Moreover, Schumm et al. modeled the turbulent heat flux with a Simple Gradient Diffusion Hypothesis (SGDH). In addition, they evaluated the turbulent Prandtl number locally with the correlation of Kays~\cite{kays1994turbulent}.

In the modeling and computation of industrial turbulent flows, RANS simulation is often preferred due to its relatively lower computational cost in comparison with the more detailed large eddy simulation (LES) and direct numerical simulation. However, even RANS simulation is unfeasible for applications that require (almost) in real time modeling or testing of a large number of different system configurations, for instance for control purposes, sensitivity analyses or uncertainty quantification studies. This has motivated the development of reduced order modeling techniques. 

Reduced Basis (RB) methods, which retain the essential physics and dynamics of a high fidelity model, have been widely used in literature for the reduced order modeling of fluid flows~\cite{rozza2007reduced,veroy2003reduced}. The POD-Galerkin approach, which is a classical RB method, falls into the category of projection-based ROMs. Other types of methods are balanced truncation~\cite{willcox2002balanced,rowley2005model} and goal--oriented ROMs~\cite{bui2007goal}. Stabile et al.~\cite{stabile2019reduced} used a different POD-Galerkin based approach for the turbulence closure, namely the variational multi-scale approach. On the other hand, Carlberg et al.~\cite{carlberg2013gnat} and Xiao et al.~\cite{xiao2013non} presented a Petrov-Galerkin projection approach for the reduced order modeling of the Navier-Stokes equations.

The POD technique was introduced by Lumley~\cite{lumley1981coherent} to study the coherent structures in experimental turbulent flows. The technique is also known as the Karhunen--Loeve expansion, principal component analysis or empirical orthogonal functions. POD is used to formulate an optimal basis spanned by modes to represent the most significant features of a dynamical system and is therefore widely used in the development of reduced order models. Nevertheless, other ROM bases methods such as the dynamic mode decomposition~\cite{schmid2010dynamic,kutz2016dynamic,tissot2014model} can also be used.

The POD-Galerkin approach has recently been used by Lorenzi et al.~\cite{lorenzi2016pod} and Hijazi et al.~\cite{hijazi2018effort,hijazi2019data} to reduce the RANS equations in a finite volume framework. Stabile et al.~\cite{stabile2019reduced} used a different POD-Galerkin based approach for the turbulence closure, namely the variational multi-scale approach. Other recent efforts that deal with POD-based ROMs using a LES approach for the turbulence modeling can be found in~\cite{borggaard2008reduced,xie2017approximate,girfoglio2019finite}. On the other hand, Carlberg et al.~\cite{carlberg2013gnat} and Xiao et al.~\cite{xiao2013non} presented a Petrov-Galerkin projection approach for the reduced order modeling of the Navier-Stokes equations.

Moreover, Georgaka et al.~\cite{georgaka2018parametric} developed a POD-Galerkin reduced order model (ROM) of weakly coupled parametric Navier-Stokes and energy equations. They also included turbulence modeling in their model~\cite{georgaka2020hybrid}. On the other hand, Vergari et al.~\cite{vergari2020reduced} and also the authors of this work~\cite{starMC2019} developed a reduced order model (ROM) of buoyancy-driven flow with the employment of the Boussinesq approximation to model buoyancy-driven flows. In this work, the POD-Galerkin reduced order modeling strategy is extended for steady-state Reynolds averaged Navier--Stokes simulations of turbulent convective buoyant flows of a low-Prandtl number fluid. 

\section{Full order turbulence model}\label{sec:FOM}
The steady-state governing equations for an incompressible Newtonian fluid, based on the low-Reynolds Launder-Sharma $k$-$\epsilon$ model~\cite{launder1974application}, for turbulent buoyancy-driven flows in the mixed convection regime are
\begin{equation} \label{eq:FOM_mat1}
\nabla \cdot \boldsymbol{U} = 0,
\end{equation} 
\vspace{-0.5cm}
\begin{equation} \label{eq:FOM_mat2}
\nabla \cdot \left( \boldsymbol{U} \otimes \boldsymbol{U}\right) = - \nabla P + \nabla \cdot \left[\nu  \left(\nabla \boldsymbol{U}+\left(\nabla \boldsymbol{U}^T\right)\right) - \overline{\boldsymbol{u}'\boldsymbol{u}'}\right] - \boldsymbol{g}\beta (\theta - \theta_{ref}),
\end{equation} 
\vspace{-0.5cm}
\begin{equation} \label{eq:FOM_mat3}
\nabla  \cdot \left( \boldsymbol{U}\theta \right) = \nabla \cdot \left(\alpha \nabla \theta  - \overline{\boldsymbol{u}'\theta'}\right),
\end{equation} 
\vspace{-0.5cm}
\begin{equation} \label{eq:FOM_mat4}
\nabla  \cdot \left( \boldsymbol{U}k \right) = \nabla \left[\left( \nu + \frac{\nu_t}{\sigma_k}\right) \nabla k \right] + P_k -D - \epsilon,
\end{equation}
\vspace{-0.5cm}
\begin{equation} \label{eq:FOM_mat5}
\nabla  \cdot \left( \boldsymbol{U}\epsilon \right) = \nabla \left[\left( \nu + \frac{\nu_t}{\sigma_{\epsilon}}\right) \nabla \epsilon \right] + \frac{\epsilon}{k}\left[C_{\epsilon 1}f_1P_k - C_{\epsilon 2}f_2\epsilon \right] + E,
\end{equation}
\noindent where $\boldsymbol{U}$, $P$, and $\theta$ are the ensemble averaged fields for velocity, kinematic pressure, which is pressure divided by the fluid density $\rho$, and temperature, respectively. $\boldsymbol{u}'$ and $\theta'$ are the turbulent fluctuating components for velocity and temperature, respectively. Equations~\ref{eq:FOM_mat1},~\ref{eq:FOM_mat2} and~\ref{eq:FOM_mat3} are the continuity, momentum and energy equations, respectively. Equation~\ref{eq:FOM_mat4} is the transport equation for turbulence kinetic energy $k$ and Equation~\ref{eq:FOM_mat5} is the transport equation for the rate of dissipation of turbulence kinetic energy $\epsilon$. Furthermore, $\nu$ is the kinematic viscosity, $\nu_t$ is the eddy viscosity and $\alpha$ is the thermal diffusivity. The buoyancy is considered by the employment of the Boussinesq approximation in the last term of Equation~\ref{eq:FOM_mat2}, where $\theta_{ref}$ is a reference temperature, $\boldsymbol{g}$ the gravitational acceleration and $\beta$ the thermal expansion coefficient. To avoid numerical issues, due to large gradients of the buoyancy force, buoyant flow solvers typically use the shifted kinematic pressure $P_{rgh}=P - \boldsymbol{g} \cdot \boldsymbol{r}$, with $\boldsymbol{r}$ the position vector, rather than the static kinematic pressure $P$. The production term of $k$ in Equations~\ref{eq:FOM_mat4} and \ref{eq:FOM_mat5} is given by
\begin{equation}
P_k  = - \overline{\boldsymbol{u}'\boldsymbol{u}'} \nabla \boldsymbol{U}.
\end{equation}
Note that with this model the effect of buoyancy is not modeled in the turbulence transport equations (Equations~\ref{eq:FOM_mat4} and~\ref{eq:FOM_mat5}) \cite{rodi1982examples, markatos1984laminar, mallinson20106} which is in accordance with the model of Schumm et al.~\cite{schumm2018investigation}. The values of the constants $\sigma_k$, $\sigma_{\epsilon}$, $C_{\epsilon1}$, $C_{\epsilon2}$ and the damping functions $f_1$ and $f_2$ are listed in Table~\ref{tab:constants}. 

As low-Reynolds turbulence models are based on damping functions and the extra source terms D and E (listed in Table~\ref{tab:constants}), which enable the integration of the turbulence transport equations up to the wall, the use of turbulence wall functions is avoided. However, two equation-based turbulence models tend to over predict the turbulence length scale in flows at adverse pressure gradients~\cite{rodi1986scrutinizing} such as those found in detachment, reattachment and impinging regions. Accordingly, Schumm et al.~\cite{schumm2018investigation} concluded in their study on turbulent flow over a backward-facing step that the turbulence near-wall length scale correction of Yap~\cite{yap1987turbulent} needs to be added as an additional source term to the right hand side of the transport equation of $\epsilon$ (Equation~\ref{eq:FOM_mat5}). This correction has the form
\begin{equation}
S_\epsilon  = 0.83 \frac{\epsilon^2}{k} \left(\frac{k^{1.5}}{\epsilon l_e} -1\right) \left( \frac{k^{1.5}}{\epsilon l_e} \right)^2,
\end{equation}
where the turbulence length scale, $l_e$, is given by
\begin{equation}
l_e = C_\mu^{-0.75}\kappa y^+,
\end{equation}
where $\kappa$ is the von Karman constant, $C_\mu$ a model constant, both listed in Table~\ref{tab:constants}, and $y^+$ is the dimensionless wall distance.

Furthermore, the unclosed terms that contain products of fluctuating values, namely the Reynolds stress term $\overline{\boldsymbol{u}'\boldsymbol{u}'}$, and the turbulence heat transfer tensor $\overline{\boldsymbol{u}'\theta'}$, need to be modeled. The Reynolds stress term is defined as
\begin{equation}\label{eq:tauT}
-(\overline{\boldsymbol{u}'\boldsymbol{u}'}) = 2\nu_t \boldsymbol{S} -\frac{2}{3}k\boldsymbol{I},
\end{equation}
where $\boldsymbol{S} = \frac{1}{2}  \left[\nabla \boldsymbol{U}+\left(\nabla \boldsymbol{U}^T\right)\right]$ is the Reynolds-averaged strain rate tensor and $\boldsymbol{I}$ is the identity tensor. The eddy viscosity, $\nu_t$, is computed by
\begin{equation}
\nu_t = C_\mu f_\mu \frac{k^2}{\epsilon}
\end{equation}
with $f_\mu$ listed in Table~\ref{tab:constants}.

The turbulence heat flux tensor is modeled with the single gradient diffusion hypothesis (SGDH) given by the turbulence thermal diffusivity, $\alpha_t$, and the mean temperature gradient as follows
\begin{equation}
\overline{\boldsymbol{u}\theta'} = -\alpha_t \nabla \theta.
\end{equation}
The SGDH expresses the turbulence thermal diffusivity as the ratio between the eddy viscosity, $\nu_t$, and the turbulence Prandtl number, Pr$_t$, as
\begin{equation}\label{eq:alpha_t}
\alpha_t = \frac{\nu_t}{\text{Pr}_t}.
\end{equation}
Typically, Pr$_t$ is around 0.9 for wall-bounded flows. Here, the local correlation of Kays~\cite{kays1994turbulent} is applied to have a good fit to DNS of both turbulent flow in ducts and the turbulent external boundary layer of fluids with 0.025 $\leq$ Pr $\leq$ 0.1 ~\cite{schumm2018investigation}. Pr$_t$ is defined as
\begin{equation}\label{eq:Prt}
\text{Pr}_t = 0.85 + \frac{0.7}{\text{Pe}_t}.
\end{equation}
where Pe$_t$ is the turbulence Peclet number, as function of the Prandtl number and the eddy viscosity divided by the viscosity~\cite{weigand1997extended}, given by
\begin{equation}\label{eq:Pet}
\text{Pe}_t = \frac{\nu_t}{\nu}\text{Pr}.
\end{equation}

\begin{table}[h!]
	\caption{Low-Reynolds Launder-Sharma $k$-$\epsilon$ model constants, damping coefficients and source terms with Re$_t = \frac{k^2}{\nu\epsilon}$}.
	\centering
	\begin{tabular}[50pt]{ccccccccccc}
		\hline \\[-1em]
		$C_\mu$ & $\sigma_k$  & $\sigma_\epsilon$   & $C_{\epsilon1}$  & $C_{\epsilon2}$  & D & E & $f_\mu$ &  $f_1$ &  $f_2$ & $\kappa$\\ \hline \\[-1em]
		0.09 & 1 & 1.3 & 1.44 & 1.92 & 2$ \nu \left( \frac{\partial \sqrt{k}}{\partial x_i}\right)^2$   & 2$ \nu \nu_t \left( \frac{\partial^2 U}{\partial x_i\partial x_i}\right)^2$  & 1-0.3$e^{-Re_t^2}$ & 1 & $e^{\frac{-3.4}{\left(1+Re_t/50\right)^2}}$  & 0.41\\ \\[-1em]\hline
	\end{tabular}
	\label{tab:constants}
\end{table}

\subsection{Flow characteristics by non-dimensional numbers}
The flow characteristics of a fluid can be expressed by non-dimensional numbers. The most relevant ones for turbulent convective buoyant flow are given and explained here.

The ratio of the inertial forces to the viscous forces within the fluid is defined as the Reynolds number (Re)
\begin{equation}\label{eq:Re}
\text{Re} = \frac{U_bh}{\nu},
\end{equation}
where $U_b$ is the bulk velocity of the fluid and $h$ is the characteristic dimension, which is taken to be the step height. At high Reynolds numbers, the flow is dominated by the inertial forces and is therefore considered turbulent, which is typically for Re $>$ 4000 in channel flows. 

The Richardson number (Ri) represents the importance of natural convection to the forced convection and is used to determine whether the flow is in the forced, mixed or natural convection regime. In this context, Ri is defined as
\begin{equation}\label{eq:Ri}
\text{Ri} = \frac{\text{Gr}}{\text{Re}^2} = \frac{g \beta h^2 q''}{\lambda U_b^2},
\end{equation}
with $Gr$ the Grashof number defined as
\begin{equation}\label{eq:Gr}
\text{Gr} = \frac{g \beta  h^4 q''}{\nu^2 \lambda},
\end{equation}
where $g$ is the acceleration due to gravity, $q''$ the applied wall heat flux and $\lambda$ is the thermal conductivity of the fluid. Typically, the flow is in the forced convection regime when Ri $<$ 0.1, in the natural convection regime when Ri $>$ 10, and in the mixed regime when 0.1 $<$ Ri $<$ 10~\cite{garbrecht2017large}. 

The Stanton number (St) is given by the ratio of the heat transferred into the fluid to the thermal capacity of the fluid itself. Here the Stanton number, as function of the heat flux, is defined as
\begin{equation}\label{eq:St}
\text{St} = \frac{q''}{\rho U_b c_p  \Delta \theta} = \frac{q'' \nu}{\rho U_b \lambda Pr \Delta \theta} ,
\end{equation}
\noindent where $\Delta\theta$ is the characteristic temperature difference and $c_p$ the specific heat of the fluid.

The skin friction coefficient (c$_f$) is a function of the shearing stress exerted by the fluid on the wall surface over which it flows
\begin{equation}\label{eq:Cf}
\text{c$_f$} = \frac{\tau_w}{0.5 \rho U_b^2},
\end{equation}
where $\tau_w$ is the wall shear stress. There is a relationship between skin friction and heat transfer for steady flows, which is known in the context of Reynolds analogy~\cite{geankoplis2003transport}. 

\section{POD-Galerkin reduced order model for buoyancy-driven turbulent flows}\label{sec:ROM}
The Proper Orthogonal Decomposition method is used to create a reduced basis space that is spanned by a number of basis functions, or so-called modes, which capture the essential dynamics of the system~\cite{rozza2007reduced,chinesta2011short,hesthaven2016certified,quarteroni2015reduced}. The RB method assumes that the full order steady-state solutions, the so-called snapshots, of the discretized RANS equations for different parameter values, $\mu$, can be expressed as a linear combination of orthonormal spatial modes multiplied by parameter-dependent coefficients. For velocity, shifted kinematic pressure and temperature the approximations are given by
\begin{equation}\label{eq:approxU}
\boldsymbol{U}(\boldsymbol{x},\mu) \approx \boldsymbol{U}_r = \sum\limits_{i=1}^{N_r} \boldsymbol{\varphi}_i(\boldsymbol{x})a_{i}(\mu),
\end{equation}
\vspace{-0.5cm}
\begin{equation}\label{eq:approxp}
P_{rgh}(\boldsymbol{x},\mu) \approx P_{rgh_r} = \sum\limits_{i=1}^{N_r} \chi_i(\boldsymbol{x})a_{i}(\mu),
\end{equation}
\vspace{-0.5cm}
\begin{equation}\label{eq:approxT}
\theta(\boldsymbol{x},\mu) \approx \theta_r = \sum\limits_{i=1}^{N_r^\theta} \psi_i(\boldsymbol{x})b_{i}(\mu),
\end{equation}
\noindent where $\boldsymbol{\varphi}_i$, $\chi_i$ and $\psi_i$ are respectively the velocity, shifted kinematic pressure and temperature modes. It is assumed that velocity and pressure share the same coefficients $a_i(\mu)$, while $b_i(\mu)$ are the corresponding coefficients for temperature~\cite{lorenzi2016pod,caiazzo2014numerical}. Therefore, only two sets of variables are necessary~\cite{vergari2020reduced}. $N_r$ is the number of velocity and shifted kinematic pressure modes and $N_r^\theta$ is the number of temperature modes. 

The above assumptions can be extended to the turbulent eddy viscosity fields, $\nu_t$, and the turbulence thermal diffusivity fields, $\alpha_t$, in the following way
\begin{equation}\label{eq:approxnut}
\nu_t(\boldsymbol{x},\mu) \approx \nu_{t_r} = \sum\limits_{i=1}^{N_r^{\nu_t}} \eta_i(\boldsymbol{x})c_{i}(\mu),
\end{equation}
\vspace{-0.5cm}
\begin{equation}\label{eq:approxalphat}
\alpha_t(\boldsymbol{x},\mu) \approx \alpha_{t_r} = \sum\limits_{i=1}^{N_r^{\alpha_t}}\zeta_i(\boldsymbol{x})d_{i}(\mu),
\end{equation}
with $N_r^{\nu_t}$ the number of eddy viscosity modes and $N_r^{\alpha_t}$ the number of turbulence thermal diffusivity modes, respectively. $\eta_i(\boldsymbol{x})$ and $\zeta_i(\boldsymbol{x})$ are the eddy viscosity and the turbulence thermal diffusivity modes, respectively, and $c_i(\mu)$ and $d_i(\mu)$ the corresponding coefficients.

The optimal POD basis space for velocity, $E^{POD}_{U}$ = span($\boldsymbol{\varphi}_1$,$\boldsymbol{\varphi}_2$, ... ,$\boldsymbol{\varphi}_{N_r}$), is constructed by minimizing the difference between the snapshots and their orthogonal projection onto the reduced basis~\cite{quarteroni2014reduced} as follows
\begin{equation} \label{eq:min_u}
E^{POD}_{U} = \textrm{arg}\underset{\boldsymbol{\varphi}_1, ... ,\boldsymbol{\varphi}_{N_r}}{\textrm{min}} \frac{1}{N_r}\sum\limits_{n=1}^{N_s} \left\Vert \boldsymbol{U}_n(\boldsymbol{x}) - \sum\limits_{i=1}^{N_r} \left( \boldsymbol{U}_n(\boldsymbol{x}), \boldsymbol{\varphi_i} (\boldsymbol{x}) \right)_{L^2(\Omega)} \boldsymbol{\varphi}_i(\boldsymbol{x})\right\Vert_{L^2(\Omega)}^2,
\end{equation}
\noindent where $N_s$ is the number of collected snapshots and $N_s$ $>$ $N_r$. The same approach can be followed for the shifted kinematic pressure to determined the subspace $E^{POD}_{P_{rgh}}$ = span($\chi_1$,$\chi_2$, ... ,$\chi_{N_r}$). The $L^2$-norm is preferred for discrete numerical schemes~\cite{busto2020pod,Stabile2017CAF} with ${\left( \cdot,\cdot\right)_{L^2(\Omega)}}$ the $L^2$-inner product of the fields over the domain $\Omega$. Furthermore, as the modes are orthonormal to each other, ${\left( \boldsymbol{\varphi}_i,\boldsymbol{\varphi}_j\right)_{L^2(\Omega)}} = \delta_{ij}$ holds, where $\delta$ is the Kronecker delta.

For temperature, the subspace $E^{POD}_{\theta}$ = span($\psi_1$,$\psi_2$, ... ,$\psi_{N_r^\theta}$) is obtained by solving a minimization problem similar to Equation~\ref{eq:min_u}. The same procedure also applies for the subspaces $E^{POD}_{\nu_t}$ = span($\eta_1$,$\eta_2$, ... ,$\eta_{N_r^{\nu_t}}$) and $E^{POD}_{\alpha_t}$ = span($\zeta_1$,$\zeta_2$, ... ,$\zeta_{N_r^{\alpha_t}}$). 

The velocity POD modes are obtained by solving Equation~\ref{eq:min_u} using the following eigenvalue problem on the correlation matrix $\boldsymbol{C}$ of the velocity snapshots~\cite{Stabile2017CAF,stabile2017CAIM,sirovich1987turbulence}
\begin{equation} \label{eq:ev} 
\boldsymbol{C}\boldsymbol{Q}=\boldsymbol{Q}\boldsymbol{\lambda},
\end{equation}
\noindent where $C_{ij}$ = ${\left( \boldsymbol{U}_i,\boldsymbol{U}_j\right)_{L^2(\Omega)}}$ for $i$,$j$ = 1, ..., $N_s$ is the velocity correlation matrix, $\boldsymbol{Q}$ is a square matrix of eigenvectors and $\boldsymbol{\lambda}$ is a diagonal matrix containing the eigenvalues.
The velocity POD modes are then constructed in the following way
\begin{equation} \label{eq:POD_u}
\boldsymbol{\varphi}_i (\boldsymbol{x}) = \frac{1}{N_s\sqrt{\lambda_i}} \sum\limits_{n=1}^{N_s} \boldsymbol{U}_n(\boldsymbol{x}) Q_{i,n}\text{\hspace{0.5cm} for \hspace{0.1cm}}i = 1,...,N_r.
\end{equation}
As the same basis for velocity and shifted kinematic pressure are used, no additional stabilization, as the supremizer or Pressure Poisson Equation approach~\cite{Stabile2017CAF,stabile2017CAIM}, is needed. For the same reason, the shifted kinematic pressure modes are constructed using the previously obtained matrix of eigenvectors $\boldsymbol{Q}$
\begin{equation} \label{eq:POD_p}
\chi_i (\boldsymbol{x}) = \frac{1}{N_s}\sqrt{\lambda_i} \sum\limits_{n=1}^{N_s} P_{rgh_n}(\boldsymbol{x}) Q_{i,n}\text{\hspace{0.5cm} for \hspace{0.1cm}}i = 1,...,N_r,
\end{equation}
where $N_s$ is the number of collected shifted kinematic pressure snapshots. The temperature, eddy viscosity and turbulence thermal diffusivity POD modes are determined by solving a similar eigenvalue problem as Equation~\ref{eq:ev}. For more details on obtaining the POD modes, the reader is referred to~\cite{vergari2020reduced,Stabile2017CAF}.

To obtain a reduced order model the POD is combined with the Galerkin projection. The momentum equations (Equation~\ref{eq:FOM_mat2}) with substitution according to Equations~\ref{eq:approxU},~\ref{eq:approxp},~\ref{eq:approxT} and~\ref{eq:approxnut} are projected onto the POD basis space of velocity, $\boldsymbol{\varphi}_i(\boldsymbol{x})$. The energy equation (Equation~\ref{eq:FOM_mat3}) with substitution according to Equations~\ref{eq:approxU},~\ref{eq:approxT} and~\ref{eq:approxalphat} is projected onto the temperature spatial basis, $\psi_i(\boldsymbol{x})$. This results in the following reduced system of Ordinary Differential Equations (ODEs)
\begin{equation}\label{eq:ROM_mom}
\boldsymbol{a}^T\boldsymbol{C_r} \boldsymbol{a} = - \boldsymbol{A_r} \boldsymbol{a} + \nu (\boldsymbol{B_r} + \boldsymbol{BT_r}) \boldsymbol{a} +  \boldsymbol{c}^T (\boldsymbol{CT1_r} + \boldsymbol{CT2_r}) \boldsymbol{a}   - \boldsymbol{H_r} \boldsymbol{b},
\end{equation}
\vspace{-0.5cm}
\begin{equation}\label{eq:ROM_energy}
\boldsymbol{a}^T\boldsymbol{Q_r} \boldsymbol{b} = \alpha \boldsymbol{Y1_r} \boldsymbol{b} + \boldsymbol{d}^T  \boldsymbol{Y2_r} \boldsymbol{b},
\end{equation}
\noindent where
\begin{equation}\label{eq:ROM_matricesB}
B_{r_{ij}}  = {\left( \boldsymbol{\varphi}_i, \nabla \cdot \nabla  \boldsymbol{\varphi}_j  \right)_{L^{2}(\Omega)}}, 
\end{equation}
\vspace{-0.7cm}
\begin{equation}\label{eq:ROM_matricesBT}
BT_{r_{ij}}  = {\left( \boldsymbol{\varphi}_i, \nabla \cdot \left(\nabla  \boldsymbol{\varphi}_j^T \right) \right)_{L^{2}(\Omega)}}, 
\end{equation}
\vspace{-0.7cm}
\begin{equation}\label{eq:ROM_matricesC}
C_{r_{ijk}}  = {\left( \boldsymbol{\varphi}_i, \nabla \cdot (\boldsymbol{\varphi}_j \otimes \boldsymbol{\varphi}_k) \right)_{L^{2}(\Omega)}}, 
\end{equation}
\vspace{-0.7cm}
\begin{equation}\label{eq:ROM_matricesCT1}
CT1_{r_{ijk}}  = {\left(\boldsymbol{\varphi}_i, \nabla \cdot \eta_j  \nabla \boldsymbol{\varphi}_k \right)_{L^{2}(\Omega)}}, 
\end{equation}
\vspace{-0.7cm}
\begin{equation}\label{eq:ROM_matricesCT2}
CT2_{r_{ijk}}  = {\left( \boldsymbol{\varphi}_i, \nabla \cdot \eta_j \left( \nabla \boldsymbol{\varphi}_k^T \right) \right)_{L^{2}(\Omega)}}, 
\end{equation}
\vspace{-0.7cm}
\begin{equation}\label{eq:ROM_matricesD}
A_{r_{ij}}  = {\left( \boldsymbol{\varphi}_i, \nabla \chi_j \right)_{L^{2}(\Omega)}}, 
\end{equation}
\vspace{-0.7cm}
\begin{equation}\label{eq:ROM_matrices3}
H_{r_{ij}}  = {\left( \boldsymbol{\varphi}_i, (\boldsymbol{g} \cdot \boldsymbol{r}) \nabla (-\beta(\psi_j -\theta_{ref})) \right)_{L^{2}(\Omega)}},     
\end{equation}
\vspace{-0.7cm}
\begin{equation}\label{eq:ROM_matricesY1}
Y1_{r_{ij}}  = {\left( \psi_i, \nabla \cdot \nabla \psi_j  \right)_{L^{2}(\Omega)}}, 
\end{equation}
\vspace{-0.7cm}
\begin{equation}\label{eq:ROM_matricesY2}
Y2_{r_{ijk}}  = {\left( \psi_i, \nabla \cdot \left(\zeta_j \nabla \psi_k \right) \right)_{L^{2}(\Omega)}}, 
\end{equation}
\vspace{-0.7cm}
\begin{equation}\label{eq:ROM_matrices5}
Q_{r_{ijk}}  ={\left( \psi_i, \nabla \cdot (\boldsymbol{\varphi}_j \otimes \psi_k) \right)_{L^{2}(\Omega)}}.
\end{equation}
The reduced matrices associated with the linear terms and the third order tensors associated with the non-linear terms of the governing equation are stored before constructing the reduced order model during a, so called, offline stage. More details on the treatment of the non-linear terms can be found in~\cite{Stabile2017CAF}. 

Note that the system of ODEs has $N_r$ + $N_r^\theta$ + $N_r^{\nu_t}$ +$ N_r^{\alpha_t}$ unknowns, but only $N_r$ + $N_r^\theta$ equations to solve. Therefore, the coefficients $c_i(\mu^*)$ and $d_i(\mu^*)$ for any new value of an input parameter $\mu^*$ are computed with a non-intrusive interpolation procedure using Radial Basis Functions (RBF), as described in~\cite{lazzaro2002radial}. Here the procedure is described for obtaining the eddy viscosity coefficients $c_i(\mu^*)$; the procedure can applied in a similar fashion to the turbulence thermal diffusivity coefficients $d_i(\mu^*)$.

The RBF approach assumes that the coefficients $c_i(\mu^*)$ can be approximated for any new value of input parameter $\mu^*$ as a linear combination of $N_r^{\nu_t}$ chosen RBF kernels $\Theta_i$~\cite{walton2013reduced} as follows
\begin{equation}\label{eq:coeff_nut}
c_i(\mu^*)= \sum_{j = 1}^{N_s^{\nu_t}} w_{ij} \Theta_{i}\left(\| \mu^* - \mu_j \|_{L^2} \right) \text{ for } i = 1,2, ... N_r^{\nu_t},
\end{equation}
where $N_s^{\nu_t}$ is the number of eddy viscosity snapshots, $\mu_j$ are the sampling points corresponding to the eddy viscosity snapshots $\nu_{t_j}$ and $w_{ij}$ are the weights that need to be determined. These weights are calculated by solving the following linear system
\begin{equation}\label{eq:coeff_nut3}
\sum_{j = 1}^{N_s^{\nu_t}} w_{ij} \Theta_{i}\left(\| \mu_k - \mu_j \|_{L^2} \right) = c_{ik}  \text{ for }  i  = 1,2, ...  N_r^{\nu_t}  \text{ and } k = 1,2, ... N_s^{\nu_t},
\end{equation}
where the output $c_{ik}$ is a set of known eddy viscosity coefficients that are calculated by projecting the eddy viscosity snapshots $\nu_{t_k}$ obtained for the parameter inputs $\mu_k$ for $k$ = 1, 2, ..., $N_s^{\nu_t}$ onto the obtained spatial eddy viscosity modes $\eta_i$ (Equation~\ref{eq:approxnut}) in the following way
\begin{equation}\label{eq:coeff_nut2}
c_{ik}  =  {\left(\nu_{t_k},\eta_i\right)_{L^{2}(\Omega)}}  \text{ for }  i  = 1,2, ...  N_r^{\nu_t}  \text{ and } k = 1,2, ... N_s^{\nu_t}.
\end{equation}
Various kernels, $\Theta_i$, can be used for the RBFs. In this work, Gaussian kernels are considered, which have a local response, meaning that their best response is in the area near the center, in contrast to multi-quadratic RBFs which have a global response. The Gaussian kernels are defined as
\begin{equation}\label{eq:RBF_Gaussian}
\Theta_i \left(\| \mu - \mu_j \|_{L^2} \right) = e^{\left( -\gamma \| \mu - \mu_j  \|_{L^2}^2 \right)} \text{ for }  i  = 1,2, ...  N_r^{\nu_t}  \text{ and } j = 1,2, ... N_s^{\nu_t},
\end{equation}
where $\gamma$ is the parameter that determines the radius of the kernel. The RBF decreases monotonically away from the center. 

Once the coefficients $c_i$ and $d_i$ for new input parameters $\mu^*$ are obtained, the set of ODEs, Equation~\ref{eq:ROM_mom} and~\ref{eq:ROM_energy}, can be solved to obtain the coefficients $\boldsymbol{a}(\mu^*)$ and $\boldsymbol{b}(\mu^*)$. For more details about using RBF in this type of reduced order modeling setting the reader is referred to~\cite{hijazi2018effort}.

The advantage of determining the coefficients with RBFs is that it is not needed to project the turbulence modeling equations (Equations~\ref{eq:FOM_mat4} and~\ref{eq:FOM_mat5}) onto the reduced basis spanned by the eddy viscosity modes. These equations are often, even when using open source codes like OpenFOAM~\cite{Jasak}, challenging to access. Also the turbulence thermal diffusivity coefficients are determined with RBFs as $\alpha_t$ is not directly proportional to $\nu_t$ due to the use of the local correlation by Kays (Equation~\ref{eq:Prt}) for the calculation of Pr$_t$ in the SGDH. 

Good initial guesses for the reduced system of ODEs (Equations~\ref{eq:ROM_mom} and \ref{eq:ROM_energy}) are obtained by projecting, respectively, the velocity and temperature snapshots for a certain parameter value $\mu$ that is close to the value of a new input parameter $\mu^*$ onto the POD basis spaces as follows
\begin{equation}\label{eq:ROM_IC_vel}
\begin{split}
a_i(\mu^*) = \left(\boldsymbol{\varphi}_i(\boldsymbol{x}),\boldsymbol{u}(\boldsymbol{x},\mu)\right)_{L^{2}(\Omega)},
\end{split}
\end{equation}
\vspace{-0.7cm}
\begin{equation}\label{eq:ROM_IC_pres}
\begin{split}
b_i(\mu^*) = \left(\psi_i(\boldsymbol{x}),\theta( \boldsymbol{x},\mu)\right)_{L^{2}(\Omega)}.
\end{split}
\end{equation}

\section{Treatment of the non-homogeneous boundary conditions} \label{sec:BCs}
The POD basis functions are a linear combination of the snapshots and so are their values at the boundaries~\cite{Lassila}. Therefore, when using a POD-based reduced order modeling technique, the non-homogeneous boundary conditions (BCs) are, in general, not satisfied by the ROM~\cite{bizon2012reduced}. Furthermore, the BCs are not explicitly present in the reduced system and therefore they cannot be controlled directly~\cite{lorenzi2016pod}. A common approach for handling the BCs at reduced order level is a penalty method~\cite{graham1999optimal1}. This method enforces explicitly the BCs in the ROM with a penalty factor. Originally, a penalty method, was proposed by Lions and Magènes~\cite{lions1973non} in the context of finite element methods. They introduced a penalty parameter to weakly impose the boundary conditions. In the POD-Galerkin reduced order modeling setting, the penalty method has been first introduced by Sirisup and Karniadakis~\cite{Sirisup} for the enforcement of boundary conditions. 
The value of the penalty factor $\tau$ is generally chosen arbitrary~\cite{bizon2012reduced}. Nevertheless, if the penalty factor tends to infinity a strong imposition of the boundary conditions would be approached and the ROM becomes ill-conditioned~\cite{lorenzi2016pod}. On the other hand, small values of the factor result in a weak imposition~\cite{Sirisup} and the method becomes numerically unstable~\cite{epshteyn2007estimation}. Moreover, the penalty factors $\tau$ should be larger than 0 in order to have an asymptotically stable solution~\cite{lorenzi2016pod}. Therefore, the penalty factor needs to be chosen above a threshold value for which the method is stable and converges~\cite{epshteyn2007estimation}. For these reasons, the (suitable range for the) penalty factor is often determined via a sensitivity study~\cite{lorenzi2016pod,graham1999optimal1,kalashnikova2012efficient}.

The Dirichlet BC for velocity, $\boldsymbol{u}_{BC}$, is implemented in the momentum equation as follows
\begin{equation} \label{eq:mom_penalty}
\nabla \cdot \left( \boldsymbol{U} \otimes \boldsymbol{U}\right) + \nabla P - \nabla \cdot \left[\nu  \left(\nabla \boldsymbol{U}+\left(\nabla \boldsymbol{U}^T\right)\right) - \overline{\boldsymbol{u}'\boldsymbol{u}'}\right] + \boldsymbol{g}\beta (\theta - \theta_{ref}) + \tau_U\Gamma_1(\boldsymbol{U} - \boldsymbol{u_{BC}}) = 0,
\end{equation} 
where $\Gamma_1$ is the relevant boundary of the domain $\Omega$ and $\tau_U$ is the penalty factor.

Vergari et al.~\cite{vergari2020reduced} extended the penalty method to Neumann BCs. In this work, the Neumann boundary condition is only applied for temperature on a boundary $\Gamma_2$ and is considered to be related to the heat flux on the boundary, $q_{BC}''$, in the following way
\begin{equation}\label{eq:HF}
\boldsymbol{n}\cdot\nabla\theta|_{\Gamma_2}= - \frac{q_{BC}''}{\lambda}.
\end{equation}
The Neumann temperature BC together with a Dirichlet temperature BC, $\theta_{BC}$ are enforced in the energy equation~\ref{eq:FOM_mat3} on respectively boundary $\Gamma_2$ and $\Gamma_3$ as follows
\begin{equation} \label{eq:energy_penalty}
\nabla  \cdot \left( \boldsymbol{U}\theta \right) - \nabla \left(\alpha \nabla \theta - \overline{\boldsymbol{u}'\theta'} \right) + \tau_{\nabla \theta}\Gamma_2\left(\boldsymbol{n} \cdot \nabla \theta + \frac{q_{BC}''}{\lambda} \right) + \tau_\theta \Gamma_3 (\theta - \theta_{BC})= 0,
\end{equation} 
where $\tau_\theta$ is the penalty factor for the Dirichlet BC and $\tau_{\nabla \theta}$ for the Neumann BC.

Substituting the approximated expansions (Equations~\ref{eq:approxU}-\ref{eq:approxalphat}) for the fields into Equations~\ref{eq:mom_penalty} and \ref{eq:energy_penalty} and applying the Galerkin projection results in the following reduced system of equations

\begin{equation}\label{eq:ROM_mom2}
\boldsymbol{a}^T\boldsymbol{C_r} \boldsymbol{a} + \boldsymbol{A_r} \boldsymbol{a} - \nu (\boldsymbol{B_r} + \boldsymbol{BT_r}) \boldsymbol{a} - \boldsymbol{c}^T (\boldsymbol{CT1_r} + \boldsymbol{CT2_r}) \boldsymbol{a} + \boldsymbol{H_r} \boldsymbol{b} + \tau_U\left(\boldsymbol{u}_{BC}\boldsymbol{N1_r} - \boldsymbol{O1_r}\boldsymbol{a}\right) =0,
\end{equation}
\vspace{-0.5cm}
\begin{equation}\label{eq:ROM_energy2}
\boldsymbol{a}^T\boldsymbol{Q_r} \boldsymbol{b} - \alpha\boldsymbol{Y_r} \boldsymbol{b} - \boldsymbol{d}^T  \boldsymbol{Y2_r} \boldsymbol{b}  + \tau_{\nabla \theta} \left(\frac{q_{BC}''}{\lambda}\boldsymbol{N2_r} - \boldsymbol{02_r}\boldsymbol{b}\right)+ \tau_\theta\left(\theta_{BC}\boldsymbol{N3_r} - \boldsymbol{O3_r}\boldsymbol{b}\right) = 0 ,
\end{equation}
\noindent where the new terms related to boundaries $\Gamma_1$, $\Gamma_2$ and $\Gamma_3$ are
\begin{equation}\label{eq:ROM_matricesN1}
N1_{r_{i}}  = {\left\langle \boldsymbol{\varphi}_i, 1  \right\rangle_{L^{2}(\Gamma_1)}}, 
\end{equation}
\vspace{-0.7cm}
\begin{equation}\label{eq:ROM_matricesN2}
N2_{r_{i}}  = {\left\langle \psi_i, 1  \right\rangle_{L^{2}(\Gamma_2)}}, 
\end{equation}
\vspace{-0.7cm}
\begin{equation}\label{eq:ROM_matricesN3}
N3_{r_{i}}  = {\left\langle \psi_i, 1  \right\rangle_{L^{2}(\Gamma_3)}}, 
\end{equation}
\vspace{-0.7cm}
\begin{equation}\label{eq:ROM_matricesO1}
O1_{r_{ij}}  = {\left\langle \boldsymbol{\varphi}_i, \boldsymbol{\varphi}_j  \right\rangle_{L^{2}(\Gamma_1)}}, 
\end{equation}
\vspace{-0.7cm}
\begin{equation}\label{eq:ROM_matricesO2}
O2_{r_{ij}}  = {\left\langle \psi_i,  \psi_j \right\rangle_{L^{2}(\Gamma_2)}}, 
\end{equation}
\vspace{-0.7cm}
\begin{equation}\label{eq:ROM_matricesO3}
O3_{r_{ij}}  = {\left\langle \psi_i,  \psi_j \right\rangle_{L^{2}(\Gamma_3)}}.
\end{equation}
\vspace{-0.7cm}

\section{Numerical set-up} \label{sec:setup}
\begin{figure}[h!]
	\centering
	\captionsetup{justification=centering}
	\includegraphics[width=16cm]{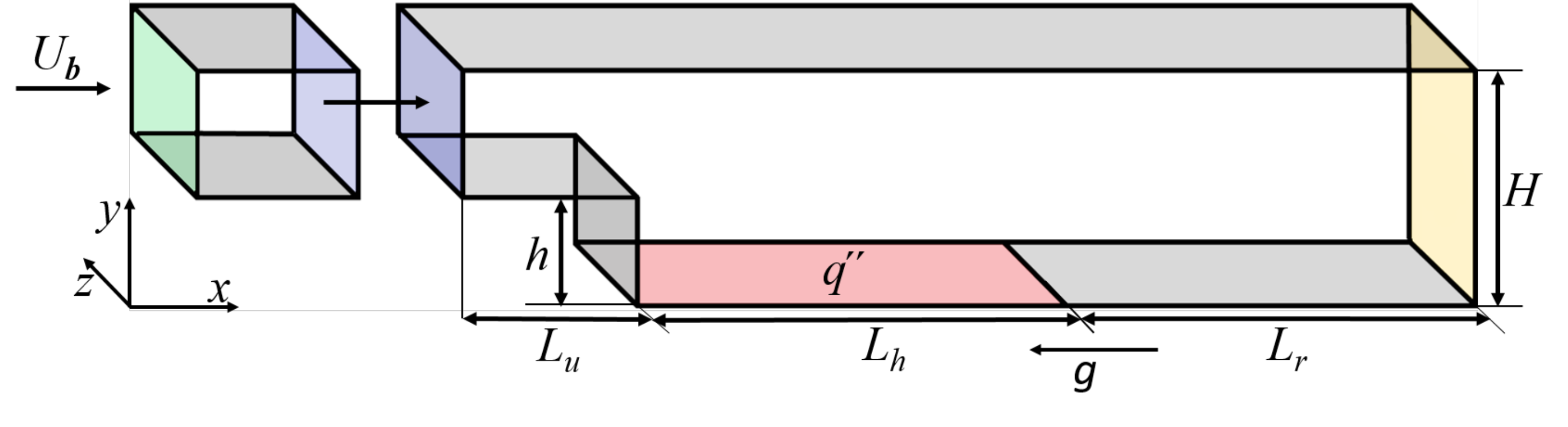}
	\caption{A sketch of the geometry of the backward-facing step and the precursor domain.}
	\label{fig:setup}
\end{figure}

In this section the numerical set-up for a backward-facing step is described. Figure~\ref{fig:setup} depicts a sketch of the geometry. The height of the step is $h$ and the channel height is $H$, which equals 2$h$. Consequently, the Expansion Ratio (ER) between inlet and outlet is ER = $H$/($H$-$h$) = 2. The inlet is located $L_u$ = 4$h$ upstream of the step. A constant heat flux is applied on the bottom wall directly downstream of the step over a length $L_h$ = 20$h$ and is referred to as "the heater". This wall is followed by an adiabatic wall of length $L_r$ = 20$h$.

\begin{figure}[h!]
	\centering
	\captionsetup{justification=centering}
	\includegraphics[width=13cm]{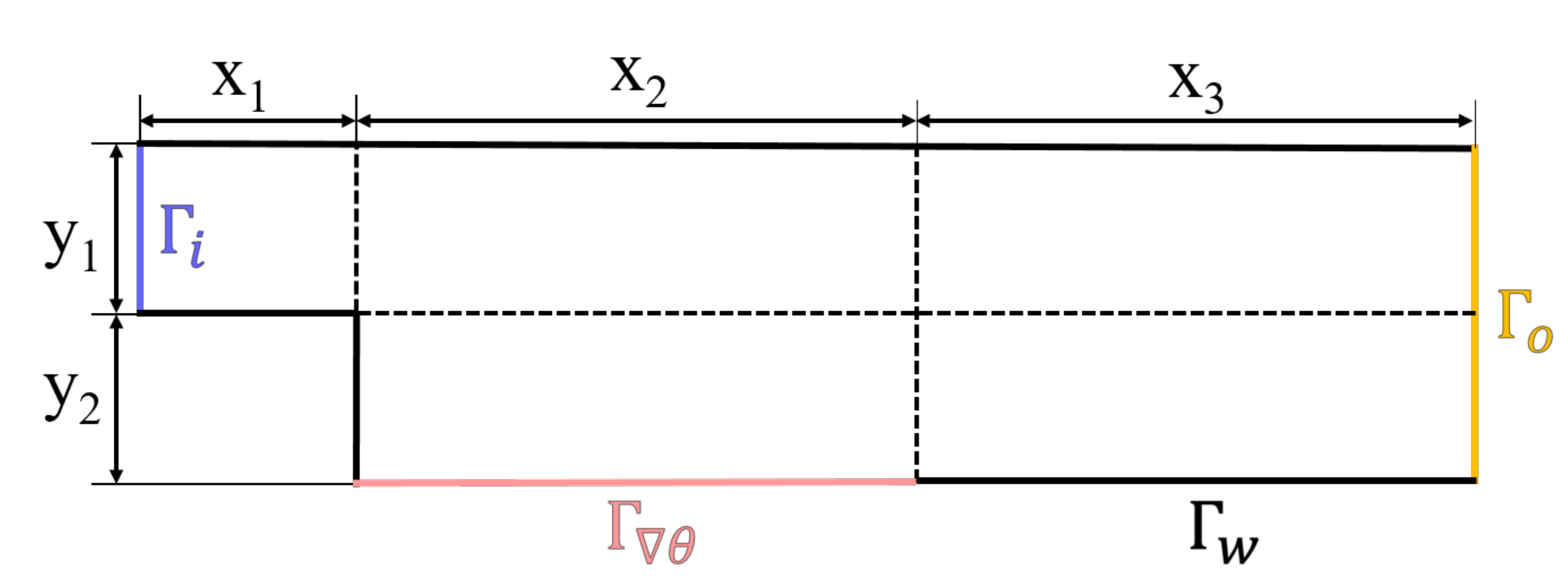}
	\caption{A 2D sketch of the geometry of the backward-facing step divided into several zones and including boundaries.}
	\label{fig:grid}
\end{figure}

A mesh is constructed in the three-dimensional domain, but can be considered to be two-dimensional as it contains only one layer of cells in the z-direction. The distribution of the cells are described in Figure~\ref{fig:grid} and in Table~\ref{tab:mesh}. These distributions are taken from the finest mesh of the grid refinement study performed by~\cite{schumm2018investigation}.  Similar to their work, the cells are clustered towards the walls and in stream-wise direction towards the end of the heater where steep changes in the velocity gradients are expected. The mesh contains a total number of 585450 hexahedra cells.  

\newpage
\begin{table}[h!]
	\caption{The number of cells, $N$, along the horizontal ($x_1$, $x_2$ and $x_3$) and vertical sides ($y_1$, $y_2$) of each zone depicted in Figure~\ref{fig:grid}.}
	\centering
	\begin{tabular}{cccccc}
		\hline
		& $N_{y1}$  & $N_{y2}$  & $N_{x1}$  & $N_{x2}$ & $N_{x3}$   \\ \hline
		Number of cells & 225 & 225 & 126 & 900 & 338 \\	
		\hline
	\end{tabular}
	\label{tab:mesh}
\end{table}

The characteristic dimension of the domain is the step height $h$ = 0.05 m. A hydrodynamic fully developed channel flow profile is applied at the inlet boundary $\Gamma_i$. This inlet velocity profile is generated via a separate simulation of an isothermal channel flow of height $h$ and length 10$h$ with the inlet bulk velocity $U_b$ = 0.1192 m/s. The flow, characterized by the Reynolds number Re = \num{e5} (Equation~\ref{eq:Re}), is considered to be fully turbulent. The hyperbolic stream-wise velocity profile at the outlet of the channel is set as the inlet velocity profile of the backward-facing step as depicted in Figure~\ref{fig:setup}. At the outlet, $\Gamma_o$, a homogeneous Neumann boundary condition is set for all variables except for pressure. Only the relative pressure is calculated and therefore it is set to 0 Pa at the outlet. At all solid walls a no-slip condition is applied and the turbulence quantities, $k$ and $\epsilon$, are set to zero. All walls except for the heated one, $\Gamma_{\nabla\theta}$, are adiabatic.

The fluid properties are taken for liquid sodium at a constant inlet temperature $\theta$ = $\theta_{in}$ = 423.15 K = 150$^\circ$C with the kinematic viscosity $\nu$ = 5.96$\cdot$\num{e-7} m$^2$/s and thermal diffusivity $\alpha$ = 6.8 $\cdot$\num{e-5} m$^2$/s, meaning that the Prandtl number, Pr, is equal to 0.0088. 

RANS simulations are performed for Richardson numbers (Equation~\ref{eq:Ri}) in a range of Ri = $\left[0.0, 0.5\right]$ with steps of 0.05, covering partly the forced- and mixed convection regime. To calculate these Richardson numbers, the thermal expansion coefficient, $\beta$ is considered to be equal to 2.5644 $\cdot$ \num{e-4} K$^{-1}$~\cite{niemann2017turbulence}. Similar to the numerical experiments done by~\cite{schumm2018investigation}, the backward-facing step is placed vertically by having the gravitational acceleration in the downward direction with $\boldsymbol{g}$ = (-9.81, 0, 0) m/s$^2$. 

The steady-state RANS equations (Equations~\ref{eq:FOM_mat1}-\ref{eq:FOM_mat5}) are discretized by the Finite Volume method with the open source C++ library OpenFOAM 6~\cite{Jasak}. The simulations are run in parallel on 8 Intel\textsuperscript{\tiny\textregistered} Xeon\textsuperscript{\tiny\textregistered} E5-2680 v3 @ 2.50GHz cores. The SIMPLE algorithm for the pressure-velocity coupling is used~\cite{ferziger2002computational} and blended schemes with an order of accuracy between one and two have been used for the spatial discretization. A solution is assumed to be converged when the scaled residuals of all variables are below \num{e-5}. 

The $y^+$ values at the heater for Ri = 0.0, 0.2, and 0.4 are compared with the values obtained by Schumm et al.~\cite{schumm2018investigation} for the same distribution of the cells in Table~\ref{tab:yplus} and similar values are observed.

\begin{table}[h!]
	\caption{$y^+$ values at the heater compared with the results of~\cite{schumm2018investigation}.}
	\centering
	\begin{tabular}{ccccccc}
		\hline
		Ri  & $y_{min}^+ $  & $y_{min}^+$~\cite{schumm2018investigation} & $y_{max}^+$  & $y_{max}^+ $~\cite{schumm2018investigation}  & $y_{avg}^+$  & $y_{avg}^+$~\cite{schumm2018investigation}   \\ \hline
		0.0 & 5.9$e^{-5}$ &2.5$e^{-3}$ &3.1$e^{-1}$ &5.6$e^{-1}$ &1.7$e^{-1}$  & 3.3$e^{-1}$ \\
		0.2 & 4.2$e^{-3}$ &1.2$e^{-3}$ &4.4$e^{-1}$ &0.9$e^{-1}$ &3.3$e^{-1}$  & 7.1$e^{-2}$ \\
		0.4 & 7.5$e^{-3}$ &1.9$e^{-3}$ &6.2$e^{-1}$ &1.1$e^{-1}$ &4.7$e^{-1}$   & 9.1$e^{-2}$ \\
		\hline
	\end{tabular}
	\label{tab:yplus}
\end{table}

The calculation of the POD modes, the Galerkin projection of the RANS solutions on the reduced subspace and the ROM simulations are carried out with ITHACA-FV~\cite{ITHACA} on a single Intel\textsuperscript{\tiny\textregistered} Xeon\textsuperscript{\tiny\textregistered} core. ITHACA-FV is a C++ library based on the Finite Volume solver OpenFOAM~\cite{Jasak}. For more details on the ITHACA-FV code, the reader is referred to~\cite{Stabile2017CAF,stabile2017CAIM,ITHACA}. 

The ROM is tested for four Richardson numbers Ri = 0.12, 0.24, 0.36 and 0.48 that are all within the aforementioned range. The ROM solutions are compared with the RANS solutions for these Richardson numbers, which are not used in the creation of the ROM basis, to check the consistency of the method.

\section{Results} \label{sec:results}
\begin{figure}[h!]
	\centering
	\includegraphics[width=1.\textwidth]{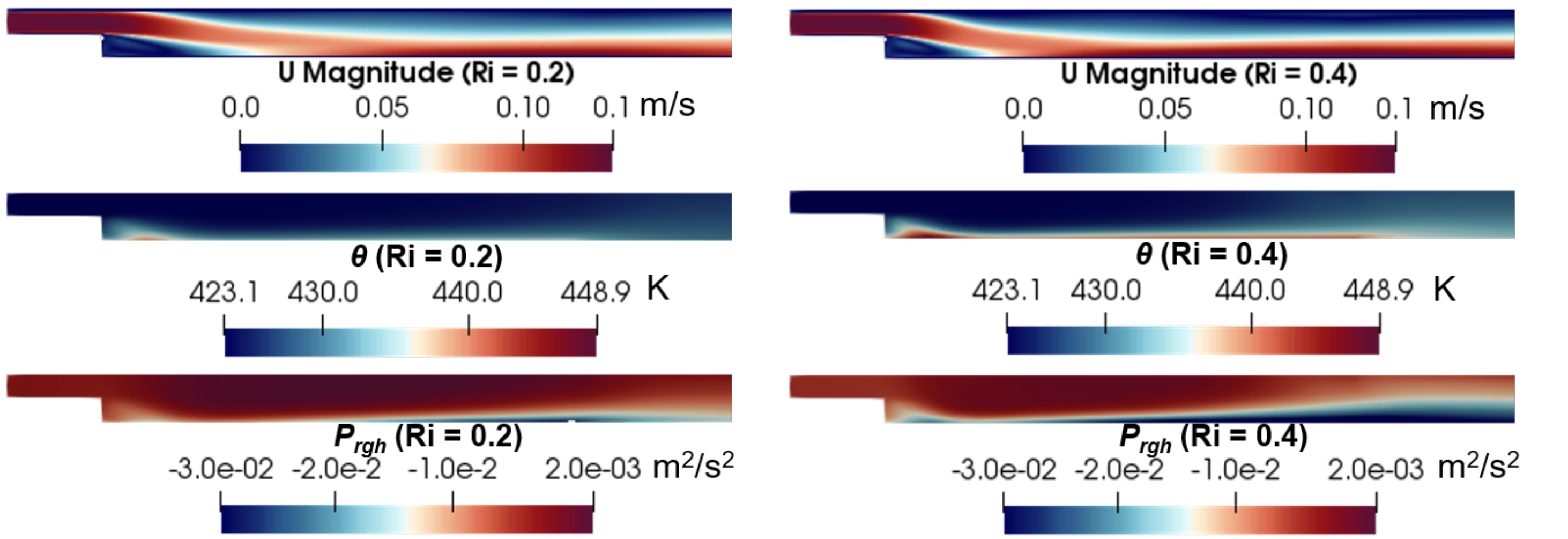}
	\caption{Velocity (top), shifted kinematic pressure (middle) and temperature fields obtained with the RANS simulations for Ri = 0.2 (left) and Ri = 0.4 (right), respectively.}
	\label{fig:FlowFields}
\end{figure}

Firstly, 10 steady-state RANS simulations are performed for the vertical backward-facing step case for Richardson numbers in the range $\left[0.05, 0.5\right]$ with steps of 0.05. The associated heat flux (Equation~\ref{eq:HF}) in the range $\left[112.5, 1125\right]$ W/m$^2$ is considered to be the corresponding varying physical parameter. The converged solutions are taken as snapshots, which are then used to create the POD basis functions. Figure~\ref{fig:FlowFields} shows the velocity magnitude, shifted kinematic pressure, and temperature fields for Ri = 0.2 (left) and Ri = 0.4 (right).

The same figure shows that the effect of buoyancy on the flow field and heat transfer is larger for higher Richardson numbers. For instance, it can be clearly seen that increasing the heat flux results in a decrease of the recirculation zone directly downstream of the step. This is also reported in~\cite{schumm2018investigation} and~\cite{niemann2016buoyancy}. 
\newpage
\subsection{Relative errors}
The approximated fields are obtained by multiplying the coefficients with the basis functions as in Equations~\ref{eq:approxU}-~\ref{eq:approxalphat}. The associated $L^2$-error between the snapshots $X_{FOM}$ and the approximated fields $X_r$, also called the (basis) projection error, is given for every parameter value, $\mu$, by 
\begin{equation}\label{eq:l2_projection}
\|\hat{e}\|_{L^2(\Omega)}(\mu) = \frac{\|X_{FOM}(\mu)-X_{r}(\mu)\|_{L^{2}(\Omega)}}{\|X_{FOM}(\mu) \|_{L^{2}(\Omega)}},
\end{equation}
\noindent where $X$ represents any field, for instance those of velocity or temperature.

\begin{figure}[h!]
	\centering
	\includegraphics[width=1.\textwidth]{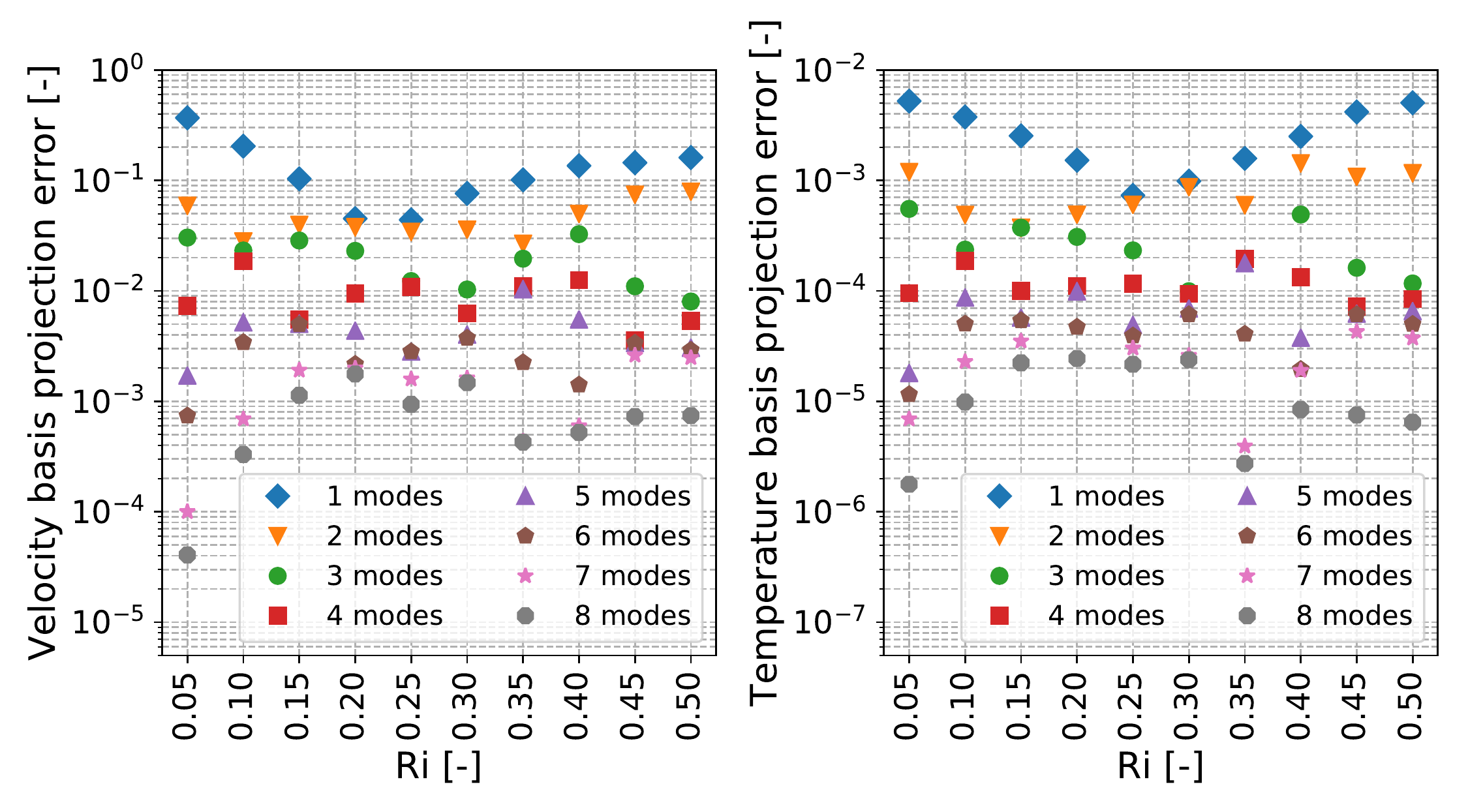}
	\caption{Relative basis projection error of all snapshots for different number of modes: (left) velocity relative error; (right) temperature relative error.}
	\label{fig:l2error_rel_non_para}
\end{figure}

Figure~\ref{fig:l2error_rel_non_para} shows the projection errors for velocity and temperature up to the first eight modes. The figure shows that for a certain parameter value the error monotonically decreases when the number of modes is increased. These errors act as a lower error bound for the reduced order model. In practice, the prediction error $\|e\|_{L^2(\Omega)}(\mu)$ for the fields obtained by solving the ROM, $X_{ROM}$, is larger than the projection error. Here the prediction error is defined as 
\begin{equation}\label{eq:l2_prediction}
\|e\|_{L^2(\Omega)}(\mu) = \frac{\|X_{FOM}(\mu)-X_{ROM}(\mu)\|_{L^{2}(\Omega)}}{\|X_{FOM}(\mu) \|_{L^{2}(\Omega)}}.
\end{equation}

In order to retain 99.99$\%$ of the energy contained in the snapshots for all physical variables 4 velocity modes, 4 shifted kinematic pressure modes, 1 temperature mode, 6 eddy viscosity modes and 6 turbulence thermal diffusivity modes are needed. Adding more modes can improve the accuracy of the ROM, but this has a detrimental effect on the computational time. Therefore, there is a trade-off between the two options. Based on Figure~\ref{fig:l2error_rel_non_para}, 5 velocity modes, 5 shifted kinematic pressure modes and 5 or 8 temperature modes are used for the construction of the ROM. The projection error is about \num{e-2} and \num{e-4} for velocity and temperature, respectively. Furthermore, 8 eddy viscosity and turbulent thermal diffusivity modes are used to accurately determine the corresponding coefficients with the RBF approach.
\newpage
\subsection{Determining the penalty factors}
\begin{figure}[h!]
	\centering
	\includegraphics[width=1.0\textwidth]{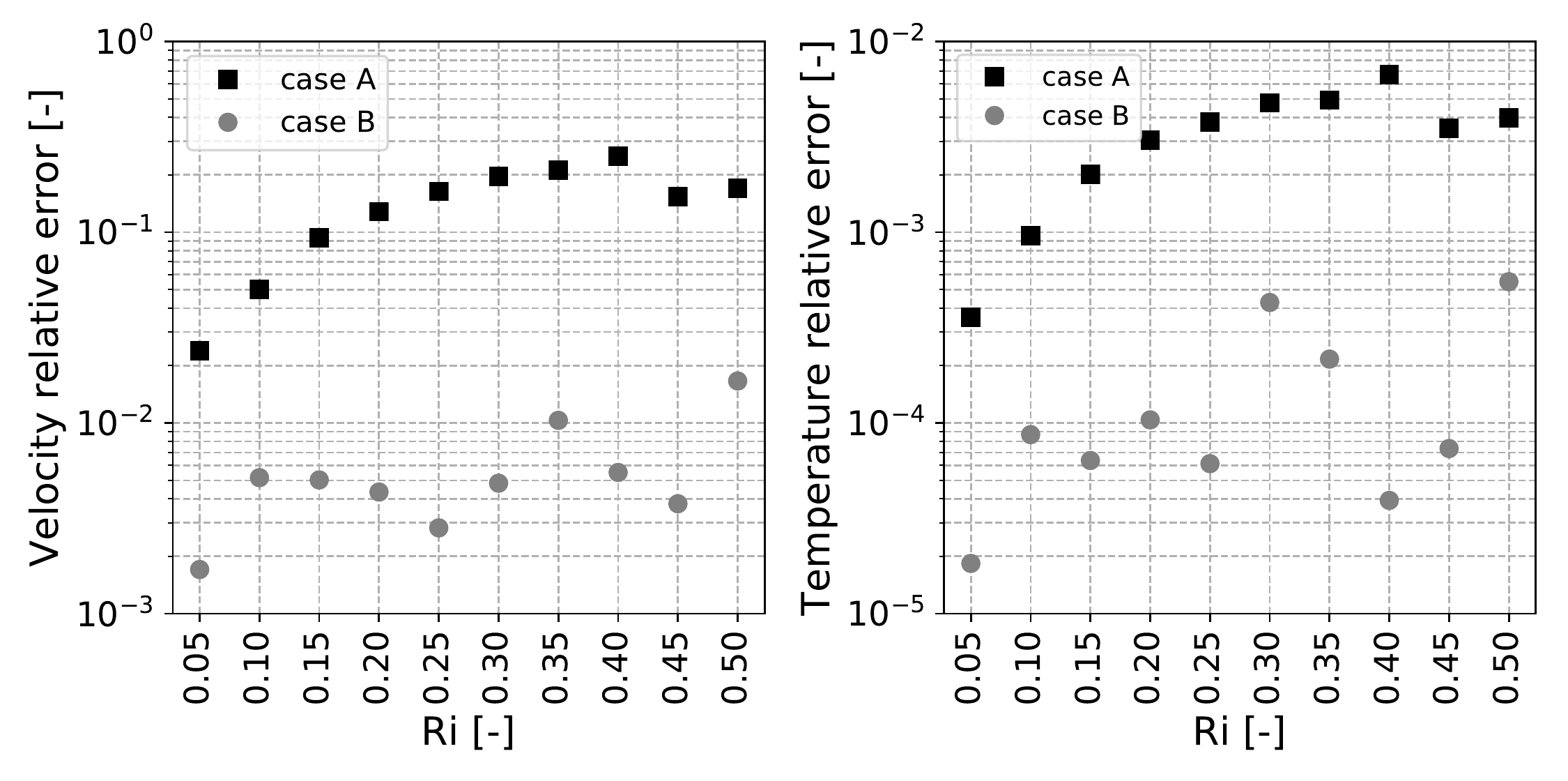}
	\caption{Relative prediction error for two sets of penalty factors and different Richardson numbers and $N_r^\theta$ = 5. Case A:  $\tau_U = 1, \tau_{\nabla\theta} = 1, \tau_\theta = 1$. Case B: $\tau_U = 10^6, \tau_{\nabla\theta} = 10^6, \tau_\theta = 10^6$. (Left) Velocity relative error; (right) temperature relative error. }
	\label{fig:l2error_non_para}
\end{figure}

The penalty factors are determined via a sensitivity study by performing multiple ROM simulations for different values of the factors. To show the effect of the penalty factors, the relative prediction errors for velocity and temperature with $N_r^\theta$ = 5 are shown in Figure~\ref{fig:l2error_non_para} for the following two cases
\begin{enumerate}
	\item[A)] $\tau_U = 1, \tau_{\nabla\theta} = 1, \tau_\theta = 1, $
	\item[B)] $\tau_U = 10^6, \tau_{\nabla\theta} = 10^6, \tau_\theta = 10^6$.
\end{enumerate}
This figure shows that the relative prediction error for both velocity and temperature improves when the penalty factors are larger. The results for other combinations of factors are not shown here as they do not lead to an overall improvement of the prediction error compared to case B. 

\newpage
\subsection{Comparison of the ROM and RANS solutions}
\begin{figure}[h!]
	\centering
	\includegraphics[width=1.0\textwidth]{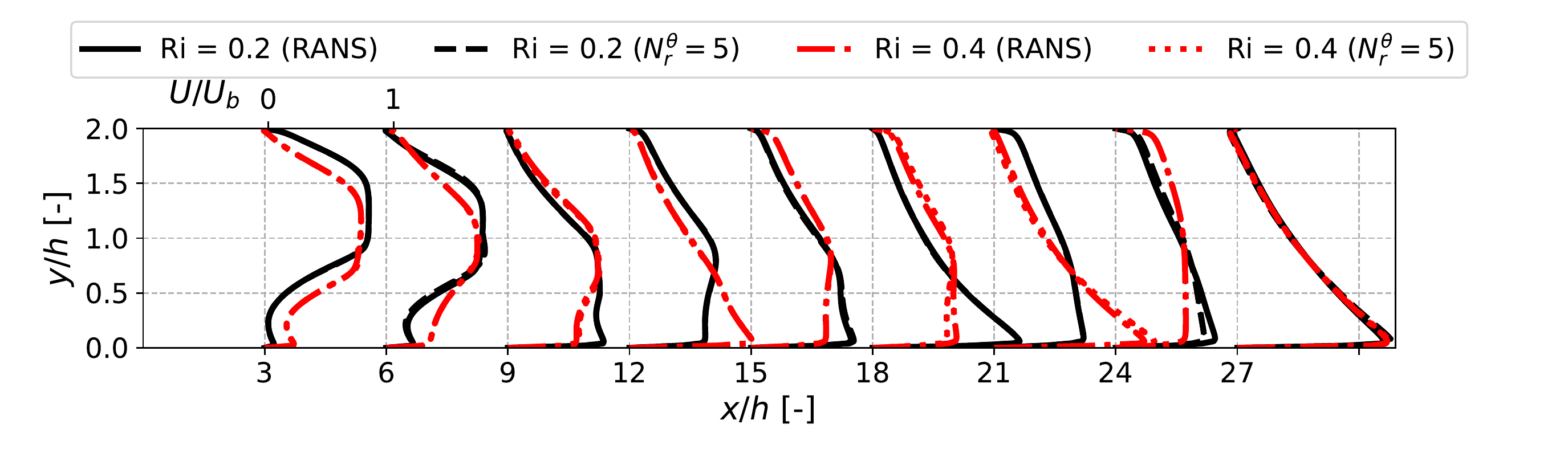}
	\caption{Profiles of the normalized stream-wise velocity component at several locations downstream of the step for Ri = 0.2 and Ri = 0.4, respectively, obtained by performing RANS and ROM simulations. The legend depicts the number of temperature modes, $N_r^\theta$, used for the ROM.}
	\label{fig:U_profiles}
\end{figure}

\begin{figure}[h!]
	\centering
	\includegraphics[width=1.0\textwidth]{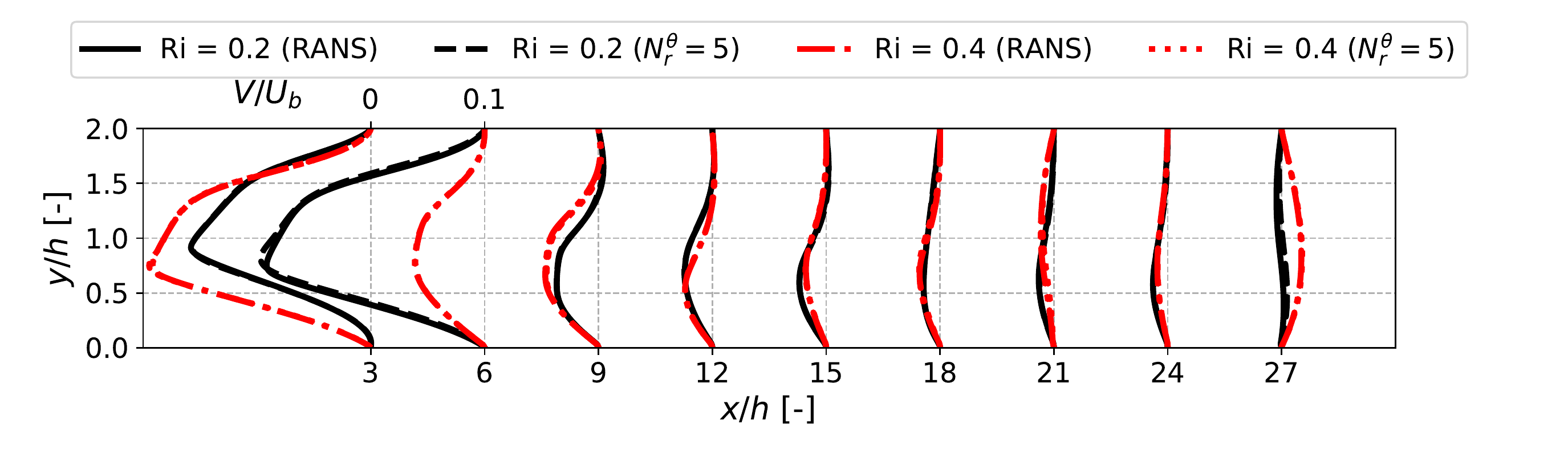}
	\caption{Profiles of the normalized wall normal velocity component at several locations downstream of the step for Ri = 0.2 and Ri = 0.4, respectively, obtained by performing RANS and ROM simulations. The legend depicts the number of temperature modes, $N_r^\theta$, used for the ROM.}
	\label{fig:V_profiles}
\end{figure}

\begin{figure}[h!]
	\centering
	\includegraphics[width=1.0\textwidth]{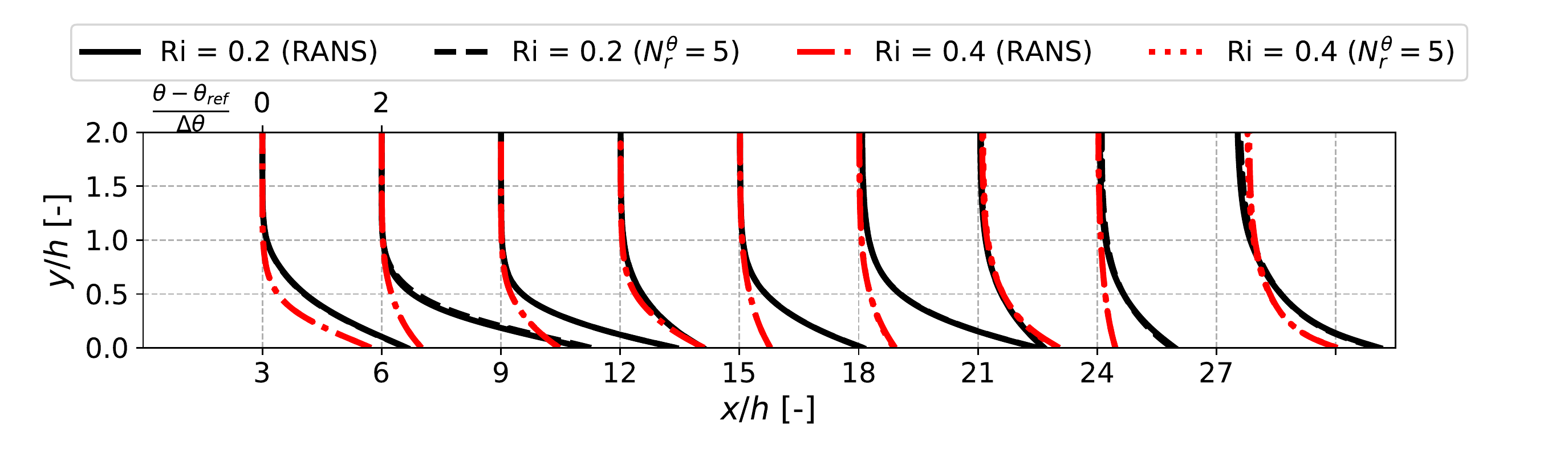}
	\caption{Normalized temperature profiles at several locations downstream of the step for Ri = 0.2 and Ri = 0.4, respectively, obtained by performing RANS and ROM simulations. The legend depicts the number of temperature modes, $N_r^\theta$, used for the ROM.}
	\label{fig:T_profiles}
\end{figure}

Reduced order simulations are performed for the same parameter values $\mu$ for which the snapshots are collected. The results of the ROM simulations are compared with the results of the RANS simulations for the cases of Ri = 0.2 and 0.4. The stream-wise velocity, the wall normal velocity and the non-dimensional temperature profiles are shown in Figure~\ref{fig:U_profiles},~\ref{fig:V_profiles} and~\ref{fig:T_profiles}, respectively. Only the results with $N_r^\theta$ = 5 are shown. Good agreement with the RANS data is found for these cases. As also observed by Schumm et al.~\cite{schumm2018investigation} the recirculation zone is reduced in its stream-wise extent (Figure~\ref{fig:U_profiles}) with increasing buoyancy. Also, the velocity profiles have their peak forming above the heater. 

Furthermore, the wall normal velocity component, shown in Figure~\ref{fig:V_profiles}, is in agreement with the results found in~\cite{schumm2018investigation}. As a positive wall normal velocity component transports momentum from the heater towards the upper wall, the temperature at the heater is lower for Ri = 0.4 compared to Ri = 0.2, as can be seen in Figure~\ref{fig:T_profiles}. Even though the velocity profiles are in good agreement with literature, the flow near the upper wall starts heating up further downstream at around $x/h$ = 27 compared to the cases studied by Schumm et al.~\cite{schumm2018investigation} where this phenomenon was already present at around $x/h$ = 15. This means that less mixing takes place in the thermal field of this study compared to their study, which can be caused by a lower shear stress. 

\begin{figure}[h!]
	\centering
	\begin{subfigure}{.51\textwidth}
		\centering
		\includegraphics[width=1.\textwidth]{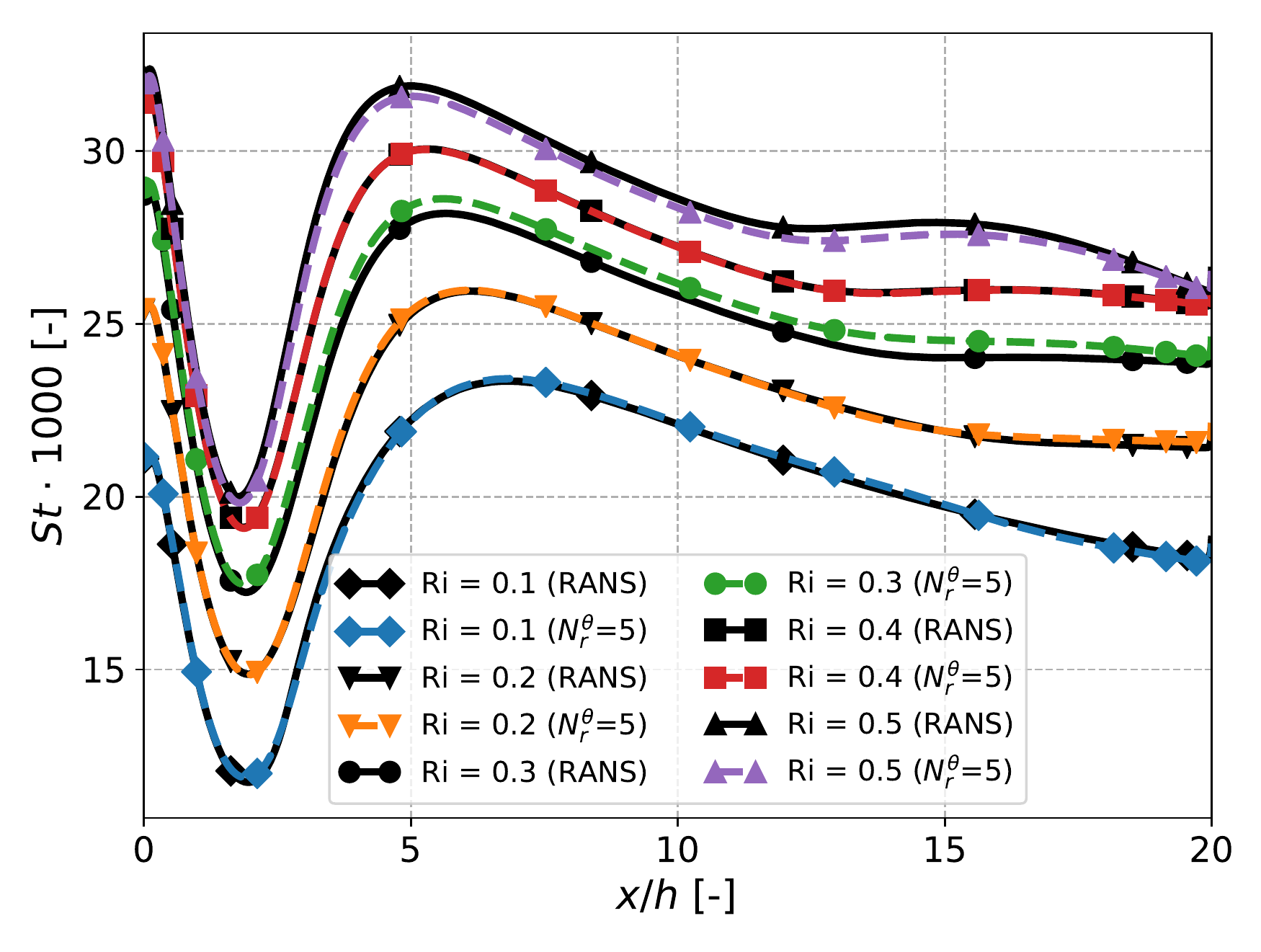}
	\end{subfigure}%
	\begin{subfigure}{.51\textwidth}
		\centering
		\includegraphics[width=1.\textwidth]{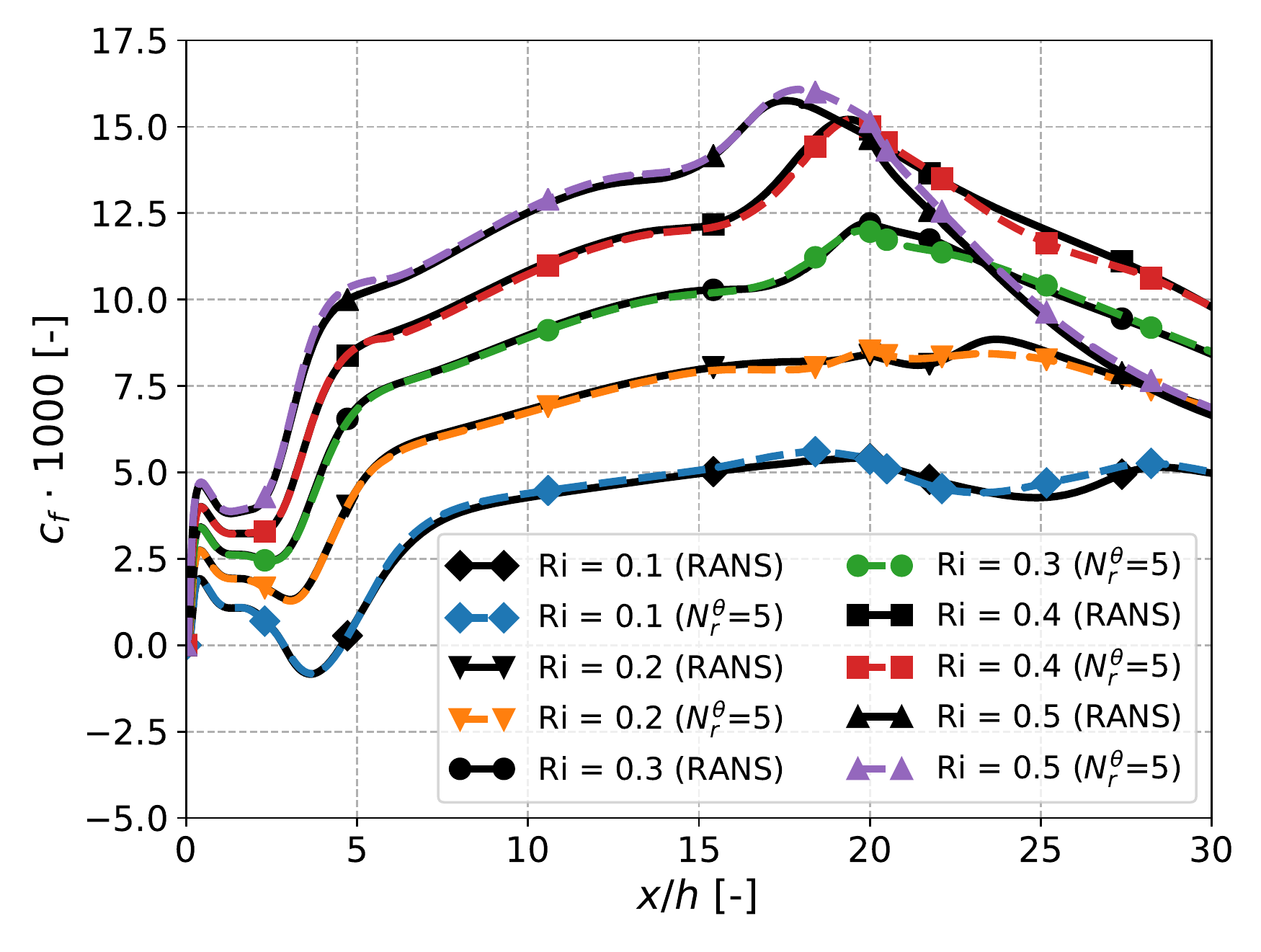}
	\end{subfigure}%
	\caption{Non-dimensional flow characteristics determined by the RANS and ROM simulations for several Richardson numbers: (left) local Stanton number at the heater; (right) skin friction distribution downstream of the backward-facing step.}
	\label{fig:non_dim_num_non_para}
\end{figure}

The local Stanton number profiles, depending on the applied heat flux, along the heater are shown in Figure~\ref{fig:non_dim_num_non_para} on the left for Ri = 0.1, 0.2, 0.3, 0.4, 0.5. The same figure on the right shows the skin friction distribution, depending on the wall shear stress, at the heater and further downstream up to $x/h$ = 30 for the same Richardson numbers on the right. Not all cases are shown for the sake of clarity. The distributions obtained by the RANS simulations are in good agreement with the literature. Furthermore, the results, and especially those of the skin friction distribution downstream of the heater, show that buoyancy has a large influence on the flow and heat transfer. This is due to the high thermal conductivity of low-Prandtl number fluids. Even though the behavior is non-linear, the reduced order model is capable of reproducing the RANS results with a good accuracy.

\subsection{Reduced order simulations for new parameter values}
\begin{figure}[h!]
	\centering
	\includegraphics[width=1.0\textwidth]{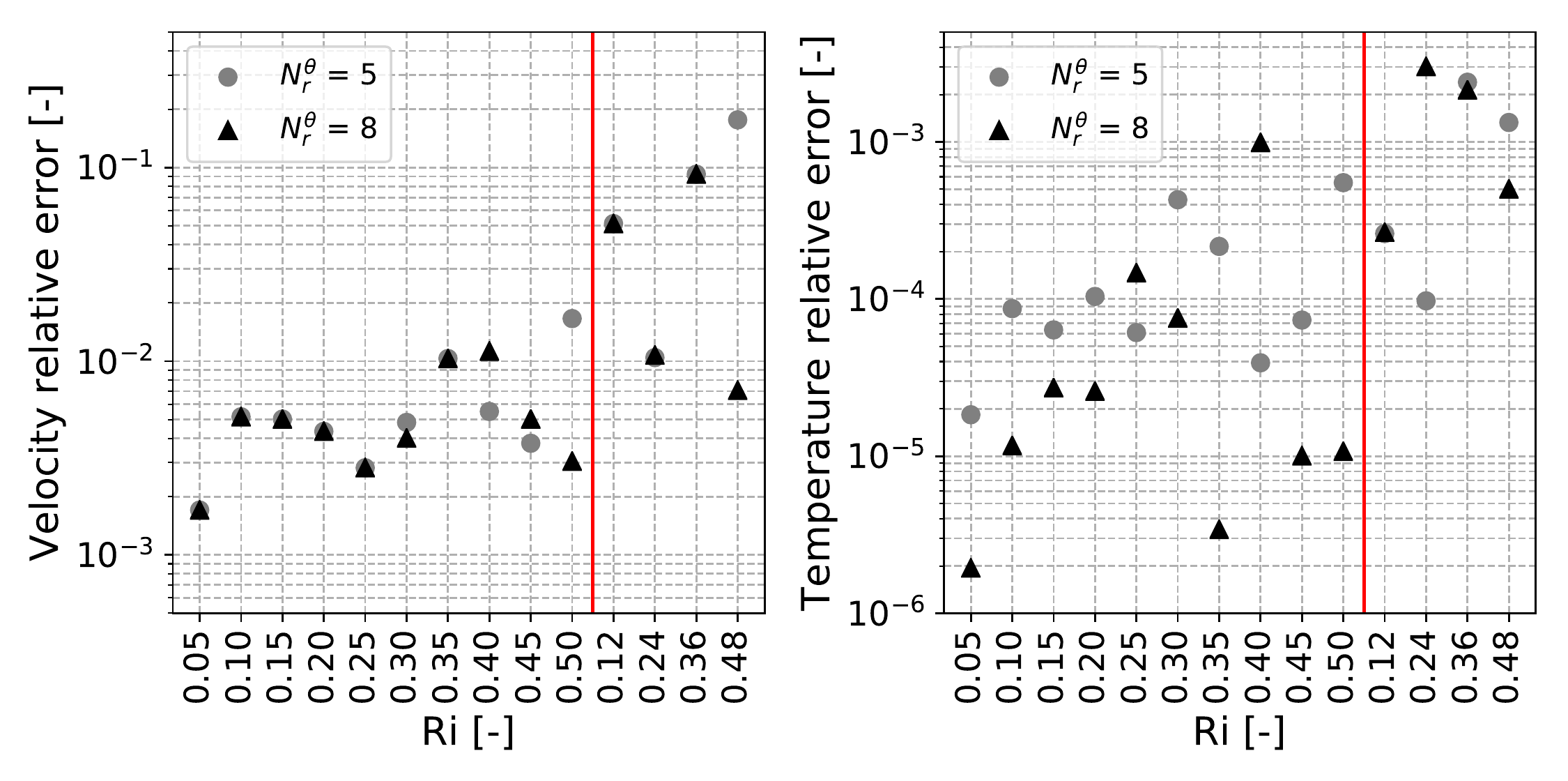}
	\caption{Relative prediction error for all Richardson numbers with $N_r^\theta$ = 5 and $N_r^\theta$ = 8 temperature modes, respectively, used for the construction of the ROM: (left) velocity relative error; (right) temperature relative error.}
	\label{fig:l2error_para}
\end{figure}

Besides the parameter for which snapshots are collected, the ROM is tested on four additional Richardson numbers, namely Ri = 0.12, 0.24, 0.36 and 0.48. Figure~\ref{fig:l2error_para} shows the prediction error for these cases with five and eight temperature modes. The figure shows that especially for Ri = 0.48 both the velocity and temperature relative error is reduced when the number of temperature modes is increased from five to eight. However, the opposite is true for Ri = 0.24. In that case, increasing the number of temperature modes has a detrimental effect on the prediction error of temperature. Therefore, five temperature modes are used further for Ri = 0.12 and 0.24, while eight temperature modes are used for Ri = 0.36 and 0.48.

\begin{figure}[h!]
	\centering
	\begin{subfigure}{.51\textwidth}
		\centering
		\includegraphics[width=1.\textwidth]{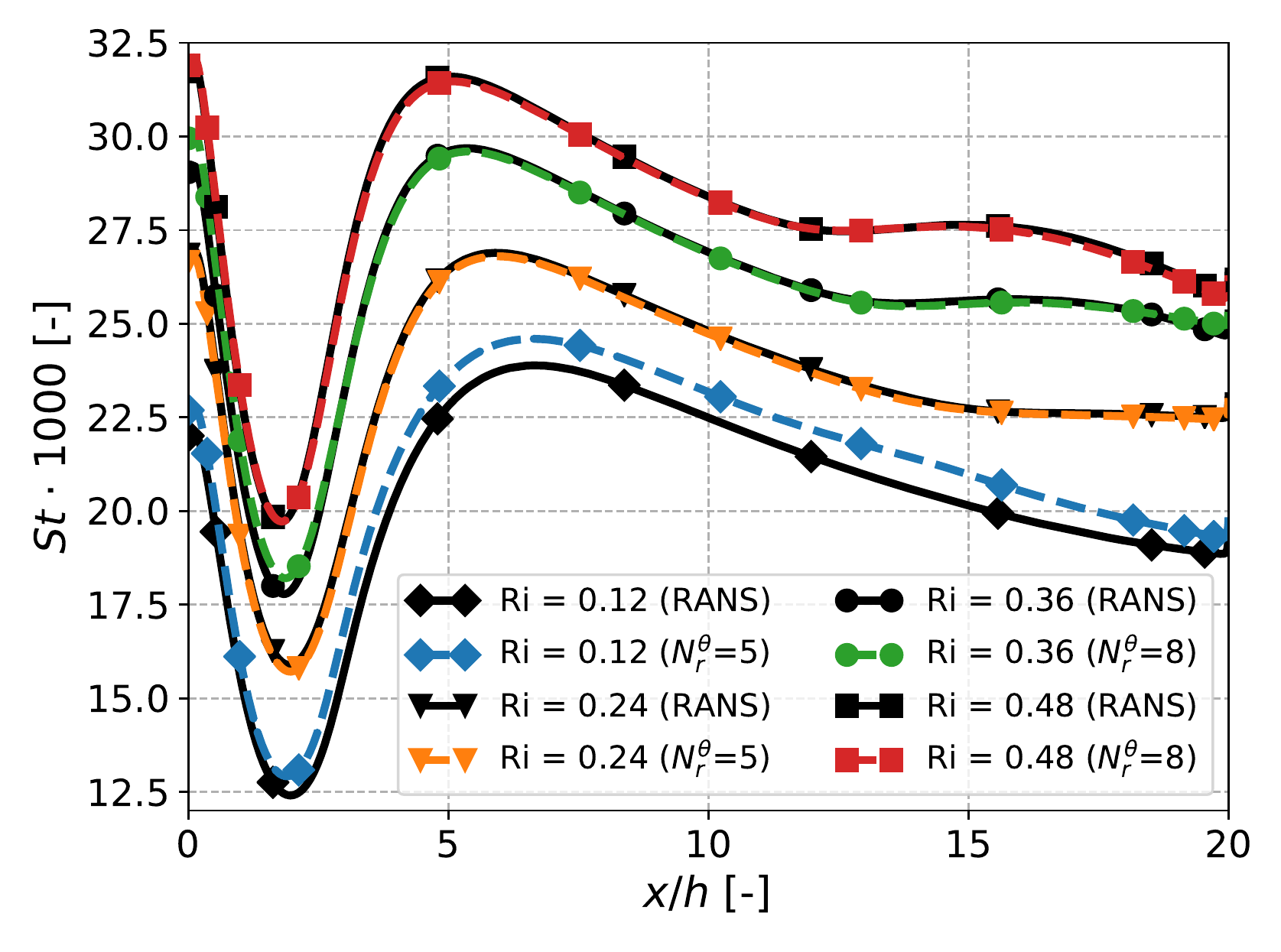}
	\end{subfigure}%
	\begin{subfigure}{.51\textwidth}
		\centering
		\includegraphics[width=1.\textwidth]{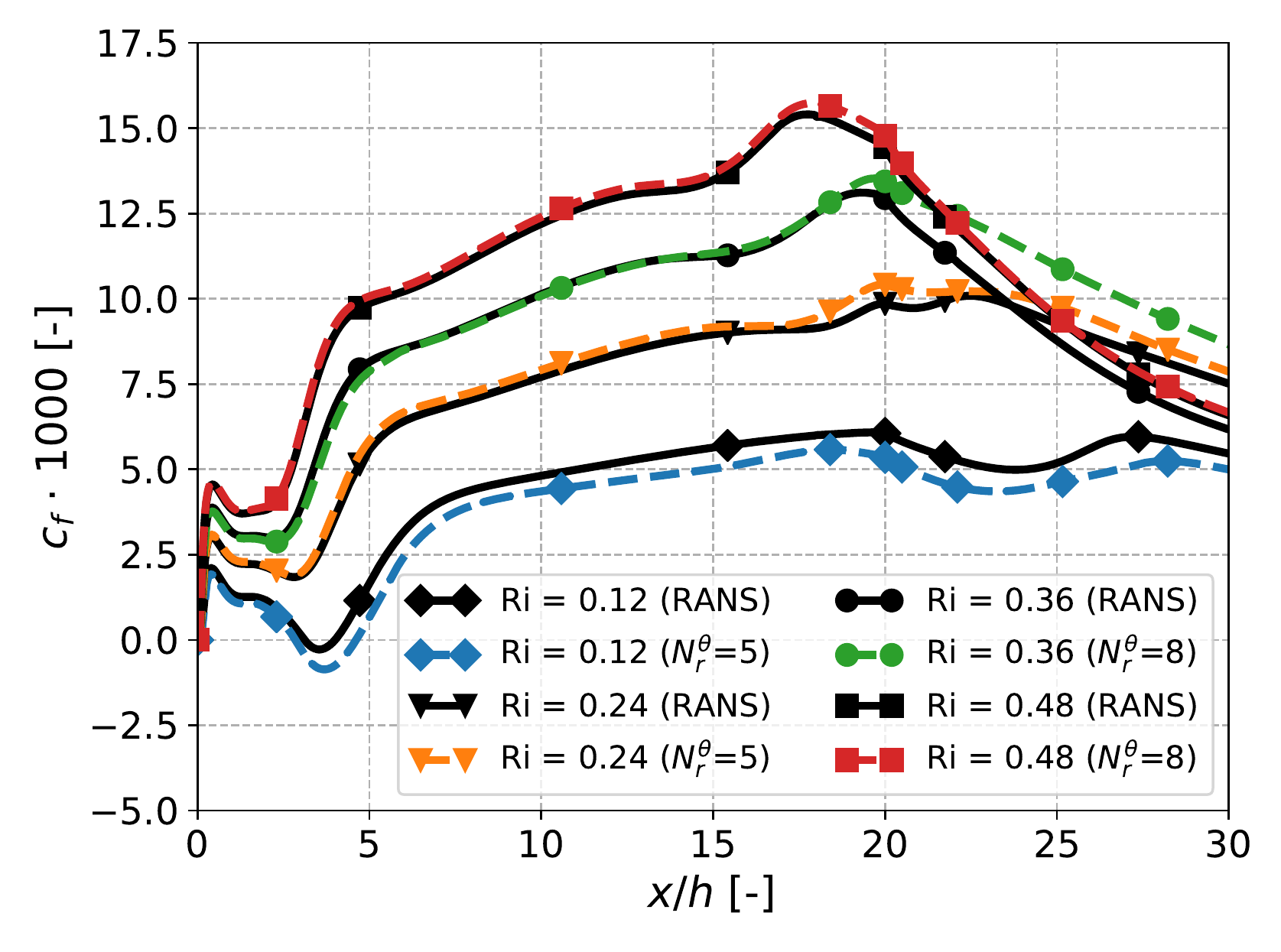}
	\end{subfigure}%
	\caption{Non-dimensional flow characteristics determined by the ROM compared with those determined by RANS simulations for several Richardson numbers: (left) local Stanton number at the heater; (right) skin friction distribution downstream of the backward-facing step.}
	\label{fig:non_dim_num_para}
\end{figure}

Figure~\ref{fig:non_dim_num_para} shows the local Stanton number and the skin friction distribution downstream of the step. The ROM results for Ri = 0.24 and 0.48 are overlapping with the distributions obtained by the RANS simulations. For Ri = 0.36 the ROM solution is accurate looking at the local Stanton number. However, the solutions for the skin friction deviates from the RANS solutions downstream of the heater. For Ri = 0.12 the ROM over-predicts the local Stanton number and under-predicts the skin friction. An attempt is made to reduce the error for this case by increasing/decreasing the number of modes of all variables and increasing/decreasing the penalty factors. However, all solutions are deviating from the ROM solution by a few percent, while for all other parameter values the deviation is less than 0.1\% in case of the local Stanton number.

\begin{figure}[h]
	\centering
	\includegraphics[width=1.0\textwidth]{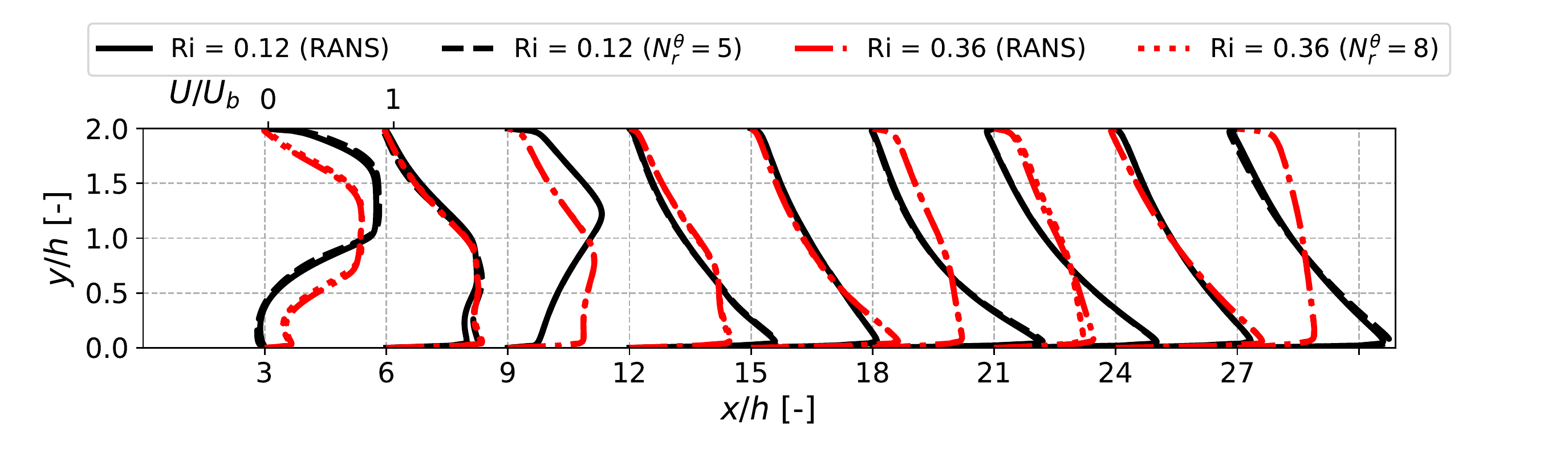}
	\caption{Profiles of the normalized stream-wise velocity component at several locations downstream of the step for Ri = 0.12 and Ri = 0.36, respectively, obtained by performing RANS and ROM simulations. The legend depicts the number of temperature modes, $N_r^\theta$, used for the ROM.}
	\label{fig:U_profiles_para_3}
\end{figure}

\begin{figure}[h]
	\centering
	\includegraphics[width=1.0\textwidth]{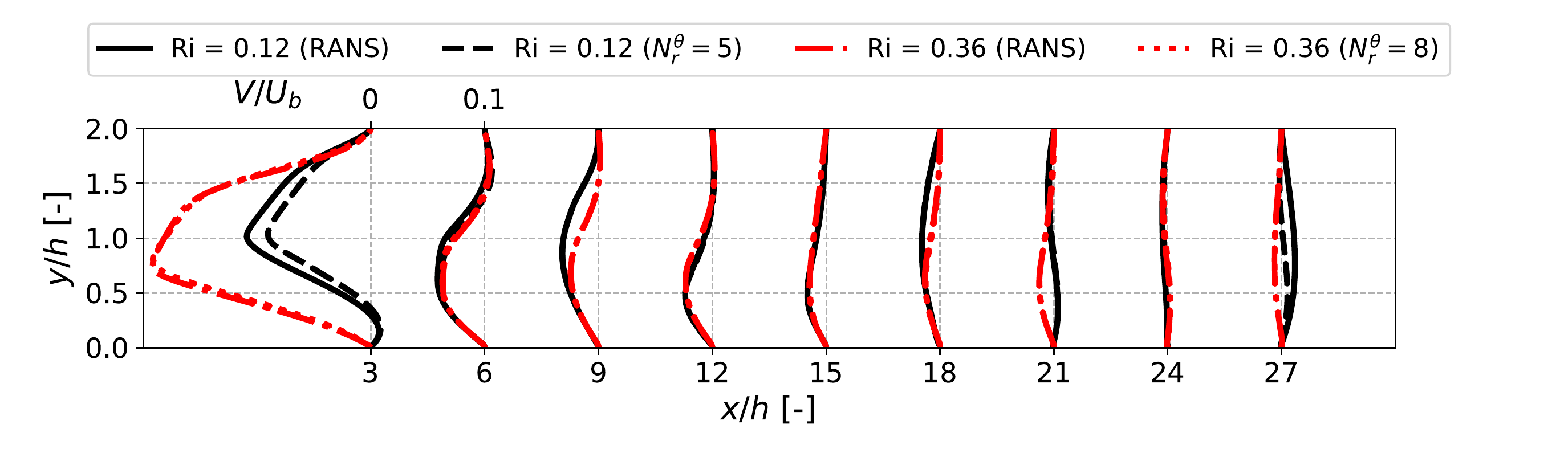}
	\caption{Profiles of the normalized wall normal velocity component at several locations downstream of the step for Ri = 0.12 and Ri = 0.36, respectively, obtained by performing RANS and ROM simulations. The legend depicts the number of temperature modes, $N_r^\theta$, used for the ROM.}
	\label{fig:V_profiles_para_3}
\end{figure}

\begin{figure}[h]
	\centering
	\includegraphics[width=1.0\textwidth]{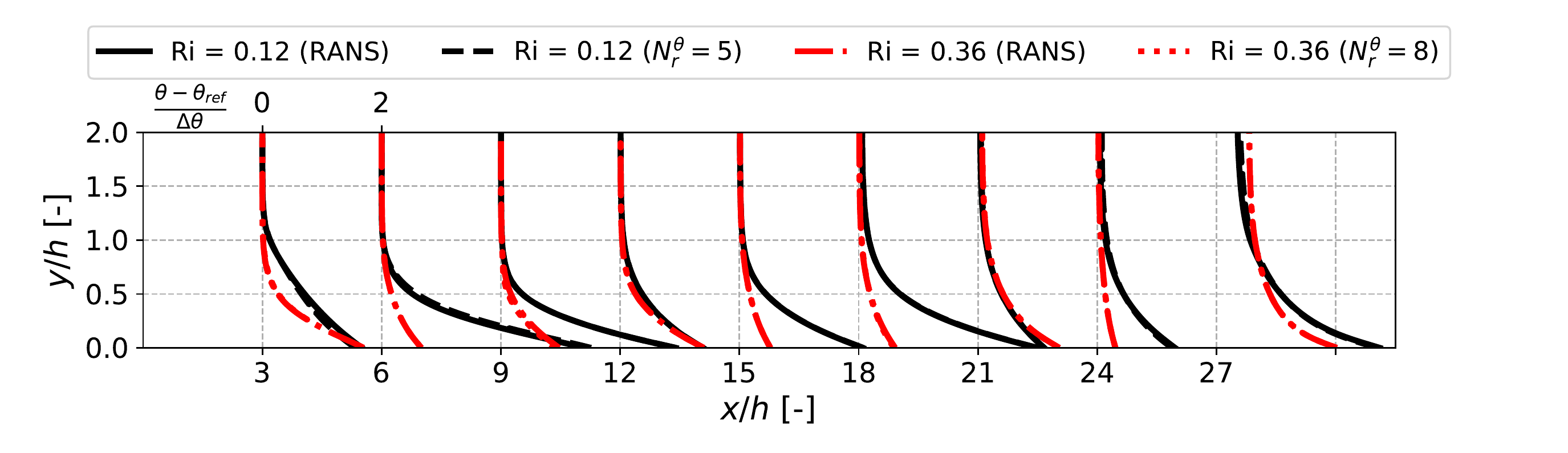}
	\caption{Normalized temperature profiles at several locations downstream of the step for Ri = 0.12 and Ri = 0.36, respectively, obtained by performing RANS and ROM simulations. The legend depicts the number of temperature modes, $N_r^\theta$, used for the ROM.}
	\label{fig:T_profiles_para_3}
\end{figure}

To see whether this affects the velocity and temperature distribution, the profiles downstream of the step are plotted in Figures~\ref{fig:U_profiles_para_3},~\ref{fig:V_profiles_para_3} and~\ref{fig:T_profiles_para_3}. Only the profiles of the wall normal velocity component at $x/h$ = 3 show a small deviation between the RANS and ROM solution for Ri = 0.12. For all other profiles, the ROM solutions are fully overlapping with the RANS solutions. 

\begin{figure}[h!]
	\begin{subfigure}{.51\linewidth}
		\centering
		\includegraphics[width=1.\linewidth]{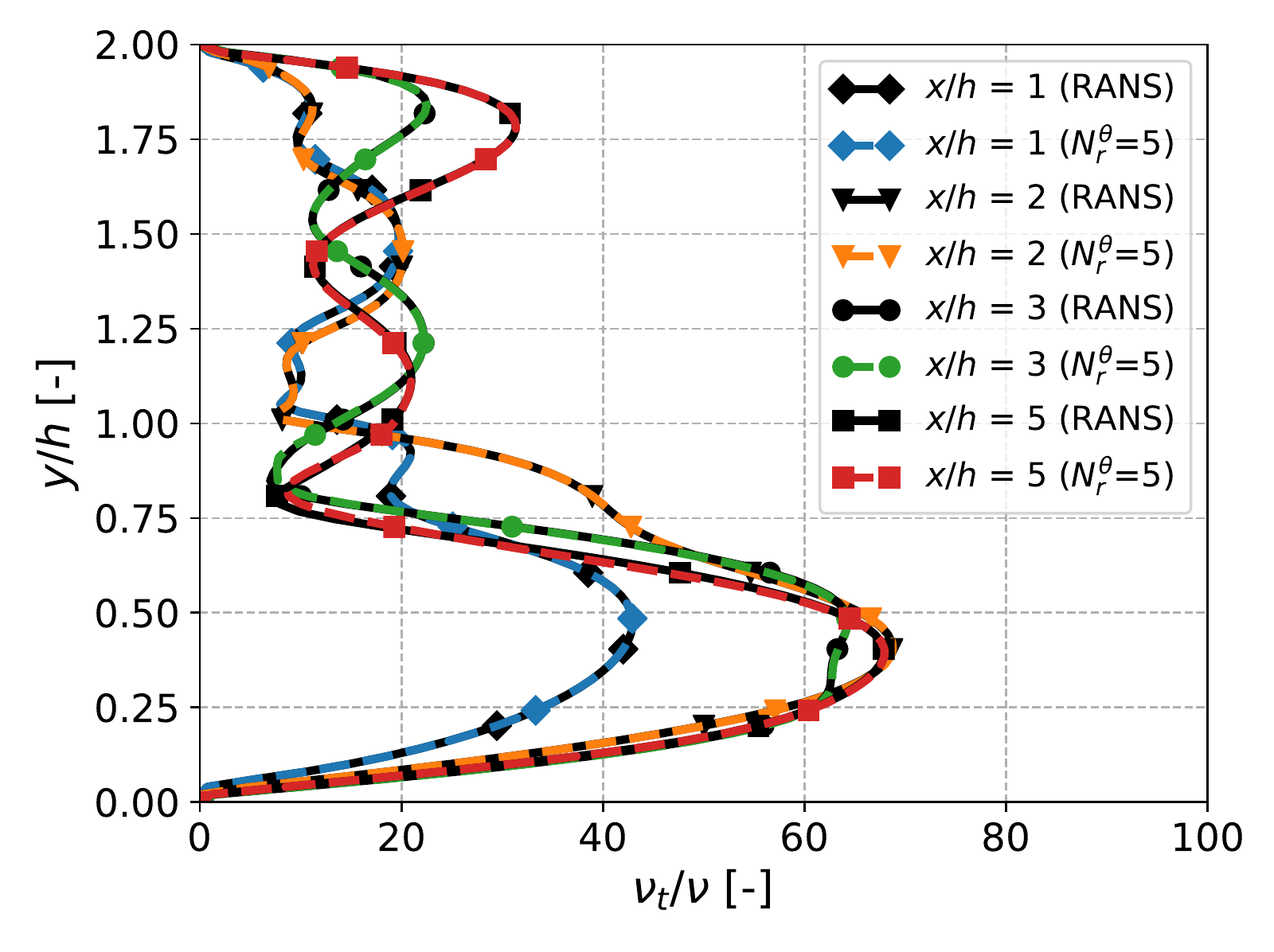}
	\end{subfigure}%
	\begin{subfigure}{.51\linewidth}
		\centering
		\includegraphics[width=1.\linewidth]{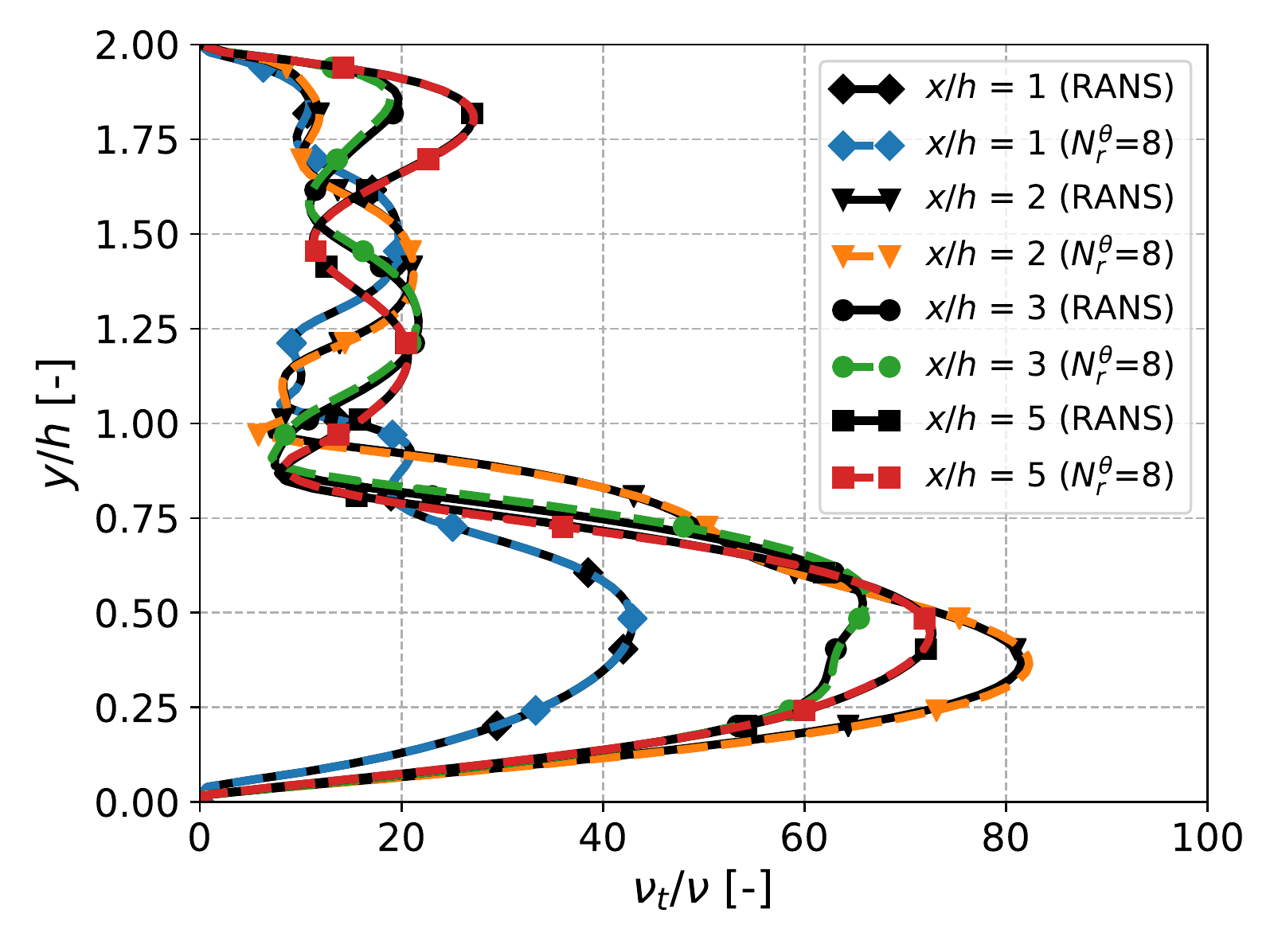}
	\end{subfigure}%
	\caption{Comparison of the ratio of eddy viscosity to kinematic viscosity at several locations downstream of the step determined by the RANS and ROM simulations for Ri = 0.12 (left) and Ri = 0.36 (right), respectively. The legend depicts the number of temperature modes, $N_r^\theta$, used for the ROM.}
	\label{fig:nu_para}
\end{figure}

\newpage
The performance of the Radial Basis Function interpolation is checked by comparing the ratio of the eddy viscosity to kinematic viscosity at several locations downstream of the step determined by the RANS and ROM simulations for Ri = 0.12 and Ri = 0.36, as shown in Figure~\ref{fig:nu_para}. Also for these fields, the ROM solutions are fully overlapping with the RANS solutions. 

\newpage

Finally, one RANS simulation takes on average 17 hours to converge on 8 Intel\textsuperscript{\tiny\textregistered} Xeon\textsuperscript{\tiny\textregistered} cores to reach a steady state solution. On the other hand, one ROM simulation takes about 1.5 seconds to converge on a single core. Therefore, the speed-up is about the order $\mathcal{O}$$\left(\num{e5}\right)$. The computational cost of the construction of the ROM is not taken into account in the calculation of the speed-up offered by the ROM itself. The whole construction of the ROM, including the collection of snapshots, calculating the POD modes and the reduced matrices, can be done in an offline phase on a high performance computing environment. For this case, the entire offline phase, which is dominated by the time it takes for the RANS simulations to converge, can be done in about 17 hours using parallel calculations on 10 (the number of snapshots) x 8 cores.

\section{DISCUSSION}\label{sec:discussion}
The results for certain Richardson numbers show that increasing the number of modes for the construction of the reduced basis space does not necessary result in a more accurate reduced order model. For instance, the relative prediction error of temperature is an order higher if eight instead of five temperature modes are used for Ri = 0.24, which can be clearly seen in Figure~\ref{fig:l2error_para}. This indicates that the ROM is not fully consistent with the high-fidelity model. 

The discrepancy between the RANS and ROM simulations can have different causes. First of all the SIMPLE algorithm is implemented in the high fidelity model, but not in the reduced order model. 

Furthermore, the turbulence transport equations for $k$ and $\epsilon$ (Equations~\ref{eq:FOM_mat4} and~\ref{eq:FOM_mat5}) are not projected on the reduced basis. Instead, the eddy viscosity and turbulence thermal diffusivity fields are approximated with an RBF interpolation approach. The advantage is that the reduced order model is independent of the turbulence model used in the RANS simulations~\cite{hijazi2018effort}. Also if the effect of buoyancy is modeled in the turbulence transport equations (Equations~\ref{eq:FOM_mat4} and~\ref{eq:FOM_mat5}), the reduced system of equations (Equations~\ref{eq:ROM_mom} and \ref{eq:ROM_energy}) does not have to be adjusted. Another option is to project the equation for the turbulence diffusivity field, Equation~\ref{eq:alpha_t} in combination with Equation~\ref{eq:Prt}, onto the reduced basis. Then the $\nu_t$ and $\alpha_t$ fields can share the same coefficients as $\alpha_t$ is depending on $\nu_t$ (Equation~\ref{eq:alpha_t}). However, this is not tested in this study. 

Moreover, some modes contain more features of the flow solution for higher Richardson numbers than others. This can be seen in Figure~\ref{fig:l2error_non_para} as the basis projection error is not the same for all Richardson numbers. Also, the projection error stagnates more or less at 7 modes, as can be seen in the same figure. This means that constructing a reduced basis with even more modes can have a detrimental effect on the ROM solution as these higher modes contain only a limited amount of physical information.

The ROM over-predicts the local Stanton number at the heater and under-predicts the skin friction for Ri = 0.12 as shown in Figure~\ref{fig:non_dim_num_para}. Neither increasing/decreasing the number of modes of all variables nor increasing/decreasing the penalty factors resulted in a lower error. In this work, the amount of RANS data is limited as only 10 snapshots are used for the construction of the reduced basis. Therefore, a possible solution is to construct a ROM with more snapshots collected in a narrow range of Richardson numbers around the parameter to be tested~\cite{Hess_CMAME_locROM}. However, this approach is time consuming as multiple local reduced order models need to be constructed.

In this study, the penalty factors are found by numerical experimentation. This can, however, be a time consuming process if the results are not satisfactory after a few tries. It remains a question how to properly select the penalty factors for the enforcement of boundary conditions in the reduced order models. This highlights one of the main drawbacks of the penalty methods, namely that the factors cannot be determined a priori~\cite{graham1999optimal1}. The authors of this work presented an iterative method to determine the factors automatically in a ROM setting in~\cite{star2019extension}, instead of performing a sensitivity study. This approach can also be used for the ROM developed in this work. 

The parametric ROM is constructed in this work to study solutions for different Richardson numbers, for which the associated heat flux is considered to be the corresponding varying physical parameter. The ROM is already set-up in such a way that it can also be used for other parameters. For instance, the constant viscosity is taken outside the reduced matrix for the diffusive term and the inlet velocity that appears as variable in the penalty term of the reduced momentum equations. Nevertheless, the ROM is not trained for these parameters even when staying within the range of Richardson numbers for which snapshots are collected. Therefore, a new ROM needs to be constructed if new or additional snapshots are needed as the POD basis functions are assumed to be based on a linear combination of the snapshots. Furthermore, geometric parametrization, like changing the height of the step, is not possible with this ROM.

The ROM can be extended to unsteady RANS simulation by incorporating a time integration method at reduced order level. However, standard POD-Galerkin ROMs tend to exhibit instabilities when an iterative algorithm for solving the non-linear implicit equations is implemented at reduced order level~\cite{baiges2014reduced,Akhtar,bergmann2009enablers,fick2017reduced}. An iterative algorithm is required due to the presence of the coupling between pressure and velocity~\cite{Stabile2017CAF}. Moreover, the snapshots do not only need to be collected in parameter space, but also at several time instances~\cite{kunisch2010optimal}.

Finally, it is known from the literature that simulating unsteady turbulent convective (buoyant) flows is challenging and RANS simulations are inaccurate for large classes of flows~\cite{drikakis2009large}. The large eddy simulation method, which is giving access to the fluctuating quantities, is often required~\cite{grotzbach1999direct,simoneau2010applications,argyropoulos2015recent}. One of the challenges of developing a LES-ROM, other than applying filtering, is the derivation of the ROM closure model to improve the accuracy and instability of the standard POD-Galerkin ROM~\cite{wang2012proper,xie2018data,osth2014need}. More research is needed to extend the current ROM for transient simulations as well as for large eddy simulation.

\section{CONCLUSIONS AND PERSPECTIVES}\label{sec:conclusion}
A Finite-Volume based POD-Galerkin reduced order modeling strategy for steady-state Reynolds averaged Navier--Stokes simulations is developed for low-Prandtl number fluid flow. Simulations are performed for sodium flow over a vertical backward-facing step with a heater placed on the wall directly downstream of the step.

The results for different Richardson numbers show that buoyancy has large influence on the flow and heat transfer, which is due to the high thermal diffusivity of low-Prandtl number fluids. Even though the behavior is non-linear, the reduced order model is capable of reproducing the RANS results with good accuracy. The prediction error between the RANS and ROM velocity fields is of the order $\mathcal{O}$(\num{e-2}) and below the order $\mathcal{O}$(\num{e-1}) for new parameter values inside the range of Richardson numbers. For temperature, the relative error is about or less than the order $\mathcal{O}$(\num{e-3}) for all parameter values. Also, the local Stanton number and skin friction distribution at the heater are qualitatively well captured. Moreover, the eddy viscosity fields are approximated well with the Radial Basis Function interpolation method.

Finally, the reduced order simulations performed on a single Intel\textsuperscript{\tiny\textregistered} Xeon\textsuperscript{\tiny\textregistered} core are about $10^5$ times faster than the RANS simulations performed on 8 cores.

For further work, the aim is to extend the reduced order model of turbulent convective buoyant flow of low-Prandtl number fluid for the parametrized unsteady RANS equations. An interesting follow-up study would be to develop a ROM for unsteady flow and heat transfer of sodium in an outlet plenum~\cite{markatos1978transient}. Furthermore, the work can be extended to large eddy simulations and compressible flows.
In addition, future work could include the use of data-driven techniques to adapt the ROM while the reduced order simulation proceeds~\cite{peherstorfer2015dynamic}. Neural Networks~\cite{frank2020machine}, instead of using Radial Basis Functions as an interpolation method, could potentially be used to approximate the eddy viscosity and thermal diffusion coefficients conducted in this work~\cite{hijazi2019data}.

\section{Acknowledgments} 
We acknowledge the support provided by the European Research Council Executive Agency by the Consolidator Grant project AROMA-CFD ``Advanced Reduced Order Methods with Applications in Computational Fluid Dynamics'' - GA 681447, H2020-ERC CoG 2015 AROMA-CFD, INdAM-GNCS 2019 projects and PRIN MIUR 2017 project (NA-FROM-PDEs).

\bibliographystyle{elsarticle-num}
\bibliography{mybibfile}  

\begin{thebibliography}{10}
\expandafter\ifx\csname url\endcsname\relax
  \def\url#1{\texttt{#1}}\fi
\expandafter\ifx\csname urlprefix\endcsname\relax\def\urlprefix{URL }\fi
\expandafter\ifx\csname href\endcsname\relax
  \def\href#1#2{#2} \def\path#1{#1}\fi

\bibitem{taler2016heat}
D.~Taler, {Heat transfer in turbulent tube flow of liquid metals}, Procedia
  Engineering 157 (2016) 148--157.
\newblock \href {http://dx.doi.org/10.1016/j.proeng.2016.08.350}
  {\path{doi:10.1016/j.proeng.2016.08.350}}.

\bibitem{niemann2017turbulence}
M.~Niemann, J.~Fr{\"o}hlich, {Turbulence budgets in buoyancy--affected vertical
  backward-facing step flow at low {P}randtl number}, Flow, Turbulence and
  Combustion 99~(3-4) (2017) 705--728.
\newblock \href {http://dx.doi.org/10.1007/s10494-017-9862-6}
  {\path{doi:10.1007/s10494-017-9862-6}}.

\bibitem{roelofs2015status}
F.~Roelofs, A.~Shams, I.~Otic, M.~B{\"o}ttcher, M.~Duponcheel, Y.~Bartosiewicz,
  D.~Lakehal, E.~Baglietto, S.~Lardeau, X.~Cheng, {Status and perspective of
  turbulence heat transfer modelling for the industrial application of liquid
  metal flows}, Nuclear Engineering and Design 290 (2015) 99--106.
\newblock \href {http://dx.doi.org/10.1016/j.nucengdes.2014.11.006}
  {\path{doi:10.1016/j.nucengdes.2014.11.006}}.

\bibitem{grotzbach2013challenges}
G.~Gr{\"o}tzbach, {Challenges in low-Prandtl number heat transfer simulation
  and modelling}, Nuclear Engineering and Design 264 (2013) 41--55.
\newblock \href {http://dx.doi.org/10.1016/j.nucengdes.2012.09.039}
  {\path{doi:10.1016/j.nucengdes.2012.09.039}}.

\bibitem{cotton1990vertical}
M.~Cotton, J.~Jackson, {Vertical tube air flows in the turbulent mixed
  convection regime calculated using a low-Reynolds-number k--${\epsilon}$
  model}, International Journal of Heat and Mass Transfer 33~(2) (1990)
  275--286.
\newblock \href {http://dx.doi.org/10.1016/0017-9310(90)90098-F}
  {\path{doi:10.1016/0017-9310(90)90098-F}}.

\bibitem{oder2019direct}
J.~Oder, A.~Shams, L.~Cizelj, I.~Tiselj, {Direct numerical simulation of
  low-Prandtl fluid flow over a confined backward facing step}, International
  Journal of Heat and Mass Transfer 142 (2019) 118436.
\newblock \href {http://dx.doi.org/10.1016/j.ijheatmasstransfer.2019.118436}
  {\path{doi:10.1016/j.ijheatmasstransfer.2019.118436}}.

\bibitem{schumm2016numerical}
T.~Schumm, B.~Frohnapfel, L.~Marocco, {Numerical simulation of the turbulent
  convective buoyant flow of sodium over a backward-facing step}, in: Journal
  of Physics: Conference Series, Vol. 745, IOP Publishing, 2016, p. 032051.
\newblock \href {http://dx.doi.org/0.1088/1742-6596/745/3/032051}
  {\path{doi:0.1088/1742-6596/745/3/032051}}.

\bibitem{schumm2018investigation}
T.~Schumm, B.~Frohnapfel, L.~Marocco, {Investigation of a turbulent convective
  buoyant flow of sodium over a backward-facing step}, Heat and Mass Transfer
  54~(8) (2018) 2533--2543.
\newblock \href {http://dx.doi.org/10.1007/s00231-017-2102-8}
  {\path{doi:10.1007/s00231-017-2102-8}}.

\bibitem{ince1989computation}
N.~Ince, B.~Launder, {On the computation of buoyancy-driven turbulent flows in
  rectangular enclosures}, International Journal of Heat and Fluid Flow 10~(2)
  (1989) 110--117.
\newblock \href {http://dx.doi.org/10.1016/0142-727X(89)90003-9}
  {\path{doi:10.1016/0142-727X(89)90003-9}}.

\bibitem{launder1974application}
B.~E. Launder, B.~Sharma, {Application of the energy-dissipation model of
  turbulence to the calculation of flow near a spinning disc}, Letters in Heat
  and Mass Transfer 1~(2) (1974) 131--137.

\bibitem{yap1987turbulent}
C.~Yap, {Turbulent heat and momentum transfer in recirculating and impinging
  flows (Ph. D. Thesis)}.

\bibitem{hsieh2004numerical}
K.~Hsieh, F.~Lien, {Numerical modeling of buoyancy-driven turbulent flows in
  enclosures}, International Journal of Heat and Fluid Flow 25~(4) (2004)
  659--670.
\newblock \href {http://dx.doi.org/10.1016/j.ijheatfluidflow.2003.11.023}
  {\path{doi:10.1016/j.ijheatfluidflow.2003.11.023}}.

\bibitem{hanjalic1996natural}
K.~Hanjali{\'c}, S.~Kenjere{\v{s}}, F.~Durst, {Natural convection in
  partitioned two-dimensional enclosures at higher Rayleigh numbers},
  International Journal of Heat and Mass Transfer 39~(7) (1996) 1407--1427.
\newblock \href {http://dx.doi.org/10.1016/0017-9310(95)00219-7}
  {\path{doi:10.1016/0017-9310(95)00219-7}}.

\bibitem{craft1996recent}
T.~Craft, N.~Ince, B.~Launder, {Recent developments in second-moment closure
  for buoyancy-affected flows}, Dynamics of Atmospheres and Oceans 23~(1-4)
  (1996) 99--114.
\newblock \href {http://dx.doi.org/10.1016/0377-0265(95)00424-6}
  {\path{doi:10.1016/0377-0265(95)00424-6}}.

\bibitem{dol1999dns}
H.~Dol, K.~Hanjali{\'c}, T.~Versteegh, {A DNS-based thermal second-moment
  closure for buoyant convection at vertical walls}, Journal of Fluid Mechanics
  391 (1999) 211--247.
\newblock \href {http://dx.doi.org/10.1017/S0022112099005327}
  {\path{doi:10.1017/S0022112099005327}}.

\bibitem{manceau2000turbulent2}
R.~Manceau, S.~Parneix, D.~Laurence, {Turbulent heat transfer predictions using
  the v2--f model on unstructured meshes}, International Journal of Heat and
  Fluid Flow 21~(3) (2000) 320--328.
\newblock \href {http://dx.doi.org/10.1016/S0142-727X(00)00016-3}
  {\path{doi:10.1016/S0142-727X(00)00016-3}}.

\bibitem{kays1994turbulent}
W.~M. Kays, {Turbulent Prandtl number--where are we?}, Journal of Heat Transfer
  116~(2) (1994) 284--295.
\newblock \href {http://dx.doi.org/10.1115/1.2911398}
  {\path{doi:10.1115/1.2911398}}.

\bibitem{rozza2007reduced}
G.~Rozza, D.~B.~P. Huynh, A.~T. Patera, Reduced basis approximation and a
  posteriori error estimation for affinely parametrized elliptic coercive
  partial differential equations, Archives of Computational Methods in
  Engineering 15~(3) (2008) 1.
\newblock \href {http://dx.doi.org/10.1007/s11831-008-9019-9}
  {\path{doi:10.1007/s11831-008-9019-9}}.

\bibitem{veroy2003reduced}
K.~Veroy, C.~Prud'Homme, A.~T. Patera, Reduced-basis approximation of the
  viscous {B}urgers equation: rigorous a posteriori error bounds, Comptes
  Rendus Mathematique 337~(9) (2003) 619--624.
\newblock \href {http://dx.doi.org/10.1016/j.crma.2003.09.023}
  {\path{doi:10.1016/j.crma.2003.09.023}}.

\bibitem{willcox2002balanced}
K.~Willcox, J.~Peraire, Balanced model reduction via the proper orthogonal
  decomposition, {AIAA Journal} 40~(11) (2002) 2323--2330.
\newblock \href {http://dx.doi.org/10.2514/2.1570} {\path{doi:10.2514/2.1570}}.

\bibitem{rowley2005model}
C.~W. Rowley, Model reduction for fluids, using balanced proper orthogonal
  decomposition, International Journal of Bifurcation and Chaos 15~(03) (2005)
  997--1013.
\newblock \href {http://dx.doi.org/10.1142/S0218127405012429}
  {\path{doi:10.1142/S0218127405012429}}.

\bibitem{bui2007goal}
T.~Bui-Thanh, K.~Willcox, O.~Ghattas, B.~van Bloemen~Waanders, Goal-oriented,
  model-constrained optimization for reduction of large-scale systems, Journal
  of Computational Physics 224~(2) (2007) 880--896.
\newblock \href {http://dx.doi.org/10.1016/j.jcp.2006.10.026}
  {\path{doi:10.1016/j.jcp.2006.10.026}}.

\bibitem{stabile2019reduced}
G.~Stabile, F.~Ballarin, G.~Zuccarino, G.~Rozza, {A reduced order variational
  multiscale approach for turbulent flows}, Advances in Computational
  Mathematics 45 (2019) 2349--2368.
\newblock \href {http://dx.doi.org/10.1007/s10444-019-09712-x}
  {\path{doi:10.1007/s10444-019-09712-x}}.

\bibitem{carlberg2013gnat}
K.~Carlberg, C.~Farhat, J.~Cortial, D.~Amsallem, The {GNAT} method for
  nonlinear model reduction: effective implementation and application to
  computational fluid dynamics and turbulent flows, Journal of Computational
  Physics 242 (2013) 623--647.
\newblock \href {http://dx.doi.org/10.1016/j.jcp.2013.02.028}
  {\path{doi:10.1016/j.jcp.2013.02.028}}.

\bibitem{xiao2013non}
D.~Xiao, F.~Fang, J.~Du, C.~Pain, I.~Navon, A.~Buchan, A.~H. Elsheikh, G.~Hu,
  {Non-linear Petrov--Galerkin methods for reduced order modelling of the
  Navier--Stokes equations using a mixed finite element pair}, Computer Methods
  In Applied Mechanics and Engineering 255 (2013) 147--157.
\newblock \href {http://dx.doi.org/10.1016/j.cma.2012.11.002}
  {\path{doi:10.1016/j.cma.2012.11.002}}.

\bibitem{lumley1981coherent}
J.~L. Lumley, Coherent structures in turbulence, in: Transition and turbulence,
  Elsevier, 1981, pp. 215--242.
\newblock \href {http://dx.doi.org/10.1016/B978-0-12-493240-1.50017-X}
  {\path{doi:10.1016/B978-0-12-493240-1.50017-X}}.

\bibitem{schmid2010dynamic}
P.~J. Schmid, {Dynamic mode decomposition of numerical and experimental data},
  Journal of fluid mechanics 656 (2010) 5--28.
\newblock \href {http://dx.doi.org/10.1017/S0022112010001217}
  {\path{doi:10.1017/S0022112010001217}}.

\bibitem{kutz2016dynamic}
J.~N. Kutz, S.~L. Brunton, B.~W. Brunton, J.~L. Proctor, Dynamic mode
  decomposition: data-driven modeling of complex systems, SIAM, 2016.
\newblock \href {http://dx.doi.org/10.1137/1.9781611974508}
  {\path{doi:10.1137/1.9781611974508}}.

\bibitem{tissot2014model}
G.~Tissot, L.~Cordier, N.~Benard, B.~R. Noack, Model reduction using dynamic
  mode decomposition, Comptes Rendus M{\'e}canique 342~(6-7) (2014) 410--416.
\newblock \href {http://dx.doi.org/10.1016/j.crme.2013.12.011}
  {\path{doi:10.1016/j.crme.2013.12.011}}.

\bibitem{lorenzi2016pod}
S.~Lorenzi, A.~Cammi, L.~Luzzi, G.~Rozza, {POD}-{G}alerkin method for finite
  volume approximation of {N}avier--{S}tokes and {RANS} equations, Computer
  Methods in Applied Mechanics and Engineering 311 (2016) 151--179.
\newblock \href {http://dx.doi.org/10.1016/j.cma.2016.08.006}
  {\path{doi:10.1016/j.cma.2016.08.006}}.

\bibitem{hijazi2018effort}
S.~Hijazi, S.~Ali, G.~Stabile, F.~Ballarin, G.~Rozza, {The effort of increasing
  Reynolds number in projection-based reduced order methods: from laminar to
  turbulent flows}, Lecture Notes in Computational Science and Engineering 136.
\newblock \href {http://dx.doi.org/10.1007/978-3-030-30705-9_22}
  {\path{doi:10.1007/978-3-030-30705-9_22}}.

\bibitem{hijazi2019data}
S.~Hijazi, G.~Stabile, A.~Mola, G.~Rozza, {Data-Driven POD–Galerkin reduced
  order model for turbulent flows}, Journal of Computational Physics 416 (2020)
  109513.
\newblock \href {http://dx.doi.org/10.1016/j.jcp.2020.109513}
  {\path{doi:10.1016/j.jcp.2020.109513}}.

\bibitem{borggaard2008reduced}
J.~Borggaard, A.~Duggleby, A.~Hay, T.~Iliescu, Z.~Wang, {Reduced-Order Modeling
  of Turbulent Flows}, in: Proceedings of MTNS, Vol. 2008, 2008.

\bibitem{xie2017approximate}
X.~Xie, D.~Wells, Z.~Wang, T.~Iliescu, {Approximate deconvolution reduced order
  modeling}, Computer Methods in Applied Mechanics and Engineering 313 (2017)
  512--534.
\newblock \href {http://dx.doi.org/10.1016/j.cma.2016.10.005}
  {\path{doi:10.1016/j.cma.2016.10.005}}.

\bibitem{girfoglio2019finite}
M.~Girfoglio, A.~Quaini, G.~Rozza, {A Finite Volume approximation of the
  Navier-Stokes equations with nonlinear filtering stabilization}, Computers \&
  Fluids 187 (2019) 27--45.
\newblock \href {http://dx.doi.org/10.1016/j.compfluid.2019.05.001}
  {\path{doi:10.1016/j.compfluid.2019.05.001}}.

\bibitem{georgaka2018parametric}
S.~Georgaka, G.~Stabile, G.~Rozza, M.~J. Bluck, Parametric {POD}-{G}alerkin
  model order reduction for unsteady-state heat transfer problems,
  Communications in Computational Physics 27~(1) (2018) 1--32.
\newblock \href {http://dx.doi.org/10.4208/cicp.oa-2018-0207}
  {\path{doi:10.4208/cicp.oa-2018-0207}}.

\bibitem{georgaka2020hybrid}
S.~Georgaka, G.~Stabile, K.~Star, G.~Rozza, M.~J. Bluck, A hybrid reduced order
  method for modelling turbulent heat transfer problems, Computers \& Fluids
  (2020) 104615\href {http://dx.doi.org/10.1016/j.compfluid.2020.104615}
  {\path{doi:10.1016/j.compfluid.2020.104615}}.

\bibitem{vergari2020reduced}
L.~Vergari, A.~Cammi, S.~Lorenzi, {Reduced order modeling approach for
  parametrized thermal-hydraulics problems: inclusion of the energy equation in
  the POD-FV-ROM method}, Progress in Nuclear Energy 118 (2020) 103071.
\newblock \href {http://dx.doi.org/10.1016/j.pnucene.2019.103071}
  {\path{doi:10.1016/j.pnucene.2019.103071}}.

\bibitem{starMC2019}
S.~K. Star, G.~Stabile, S.~Georgaka, F.~Belloni, G.~Rozza, J.~Degroote,
  {POD}-{G}alerkin reduced order model of the {B}oussinesq approximation for
  buoyancy-driven enclosed flows, MC2019, ANS, Portland (OR), USA, 2019, pp.
  2452--2461.

\bibitem{rodi1982examples}
W.~Rodi, Examples of turbulence models for incompressible flows, {AIAA Journal}
  20~(7) (1982) 872--879.

\bibitem{markatos1984laminar}
N.~C. Markatos, K.~Pericleous, Laminar and turbulent natural convection in an
  enclosed cavity, International Journal of Heat and Mass Transfer 27~(5)
  (1984) 755--772.
\newblock \href {http://dx.doi.org/10.1016/0017-9310(84)90145-5}
  {\path{doi:10.1016/0017-9310(84)90145-5}}.

\bibitem{mallinson20106}
G.~D. Mallinson, S.~E. Norris, {Fundamentals of Computational Fluid Dynamics},
  Mathematical Modeling of Food Processing (2010) 125.

\bibitem{rodi1986scrutinizing}
W.~Rodi, G.~Scheuerer, {Scrutinizing the k-$\varepsilon$ turbulence model under
  adverse pressure gradient conditions}, Journal of Fluids Engineering 108~(2)
  (1986) 174--179.
\newblock \href {http://dx.doi.org/10.1115/1.3242559}
  {\path{doi:10.1115/1.3242559}}.

\bibitem{weigand1997extended}
B.~Weigand, J.~Ferguson, M.~Crawford, {An extended Kays and Crawford turbulent
  Prandtl number model}, International Journal of Heat and Mass Transfer
  40~(17) (1997) 4191--4196.
\newblock \href {http://dx.doi.org/10.1016/S0017-9310(97)00084-7}
  {\path{doi:10.1016/S0017-9310(97)00084-7}}.

\bibitem{garbrecht2017large}
O.~Garbrecht, S.~Kabelac, R.~Kneer, {Large eddy simulation of three-dimensional
  mixed convection on a vertical plate}, Tech. rep., Lehrstuhl f{\"u}r
  W{\"a}rme-und Stoff{\"u}bertragung (2017).
\newblock \href {http://dx.doi.org/10.18154/RWTH-2018-221554}
  {\path{doi:10.18154/RWTH-2018-221554}}.

\bibitem{geankoplis2003transport}
C.~J. Geankoplis, {Transport processes and separation process principles
  (includes unit operations)}, Prentice Hall Professional Technical Reference,
  2003.

\bibitem{chinesta2011short}
F.~Chinesta, P.~Ladeveze, E.~Cueto, A short review on model order reduction
  based on proper generalized decomposition, Archives of Computational Methods
  in Engineering 18~(4) (2011) 395.
\newblock \href {http://dx.doi.org/10.1007/s11831-011-9064-7}
  {\path{doi:10.1007/s11831-011-9064-7}}.

\bibitem{hesthaven2016certified}
J.~Hesthaven, G.~Rozza, B.~Stamm, Certified reduced basis methods for
  parametrized partial differential equations, Springer International
  Publishing, 2016.
\newblock \href {http://dx.doi.org/10.1007/978-3-319-22470-1}
  {\path{doi:10.1007/978-3-319-22470-1}}.

\bibitem{quarteroni2015reduced}
A.~Quarteroni, A.~Manzoni, F.~Negri, Reduced basis methods for partial
  differential equations: an introduction, Vol.~92, Springer, 2015.
\newblock \href {http://dx.doi.org/10.1007/978-3-319-15431-2}
  {\path{doi:10.1007/978-3-319-15431-2}}.

\bibitem{caiazzo2014numerical}
A.~Caiazzo, T.~Iliescu, V.~John, S.~Schyschlowa, A numerical investigation of
  velocity--pressure reduced order models for incompressible flows, Journal of
  Computational Physics 259 (2014) 598--616.
\newblock \href {http://dx.doi.org/10.1016/j.jcp.2013.12.004}
  {\path{doi:10.1016/j.jcp.2013.12.004}}.

\bibitem{quarteroni2014reduced}
A.~Quarteroni, G.~Rozza, Reduced order methods for modeling and computational
  reduction, Vol.~9, Berlin: Springer, 2014.
\newblock \href {http://dx.doi.org/10.1007/978-3-319-02090-7}
  {\path{doi:10.1007/978-3-319-02090-7}}.

\bibitem{busto2020pod}
S.~Busto, G.~Stabile, G.~Rozza, M.~E. V{\'a}zquez-Cend{\'o}n, {POD--Galerkin
  reduced order methods for combined Navier--Stokes transport equations based
  on a hybrid FV-FE solver}, Computers \& Mathematics with Applications 79~(2)
  (2020) 256--273.
\newblock \href {http://dx.doi.org/10.1016/j.camwa.2019.06.026}
  {\path{doi:10.1016/j.camwa.2019.06.026}}.

\bibitem{Stabile2017CAF}
G.~Stabile, G.~Rozza, {Finite volume POD--Galerkin stabilised reduced order
  methods for the parametrised incompressible Navier--Stokes equations},
  Computers \& Fluids 173 (2018) 273--284.
\newblock \href {http://dx.doi.org/10.1016/j.compfluid.2018.01.035}
  {\path{doi:10.1016/j.compfluid.2018.01.035}}.

\bibitem{stabile2017CAIM}
G.~Stabile, S.~Hijazi, A.~Mola, S.~Lorenzi, G.~Rozza, {{POD}-{G}alerkin reduced
  order methods for {CFD} using Finite Volume Discretisation: vortex shedding
  around a circular cylinder}, Communications in Applied and Industrial
  Mathematics 8~(1) (2017) 210--236.
\newblock \href {http://dx.doi.org/10.1515/caim-2017-0011}
  {\path{doi:10.1515/caim-2017-0011}}.

\bibitem{sirovich1987turbulence}
L.~Sirovich, Turbulence and the dynamics of coherent structures. {I}. coherent
  structures, Quarterly of Applied Mathematics 45~(3) (1987) 561--571.
\newblock \href {http://dx.doi.org/10.1090/qam/910462}
  {\path{doi:10.1090/qam/910462}}.

\bibitem{lazzaro2002radial}
D.~Lazzaro, L.~B. Montefusco, {Radial basis functions for the multivariate
  interpolation of large scattered data sets}, Journal of Computational and
  Applied Mathematics 140~(1-2) (2002) 521--536.
\newblock \href {http://dx.doi.org/10.1016/S0377-0427(01)00485-X}
  {\path{doi:10.1016/S0377-0427(01)00485-X}}.

\bibitem{walton2013reduced}
S.~Walton, O.~Hassan, K.~Morgan, Reduced order modelling for unsteady fluid
  flow using proper orthogonal decomposition and radial basis functions,
  Applied Mathematical Modelling 37~(20-21) (2013) 8930--8945.
\newblock \href {http://dx.doi.org/10.1016/j.apm.2013.04.025}
  {\path{doi:10.1016/j.apm.2013.04.025}}.

\bibitem{Jasak}
H.~Jasak, A.~Jemcov, {\^Z}.~Tukovi{\'c}, Open{FOAM}: A {C}++ library for
  complex physics simulations, Intl. workshop on coupled methods in numerical
  dynamics, Croatia, 2007.

\bibitem{Lassila}
T.~Lassila, A.~Manzoni, A.~Quarteroni, G.~Rozza, {Model Order Reduction in
  Fluid Dynamics: Challenges and Perspectives}, in: {Reduced Order Methods for
  Modeling and Computational Reduction}, Springer International Publishing,
  2014, pp. 235--273.
\newblock \href {http://dx.doi.org/10.1007/978-3-319-02090-7_9}
  {\path{doi:10.1007/978-3-319-02090-7_9}}.

\bibitem{bizon2012reduced}
K.~Bizon, G.~Continillo, {Reduced order modelling of chemical reactors with
  recycle by means of POD-penalty method}, Computers \& chemical engineering 39
  (2012) 22--32.

\bibitem{graham1999optimal1}
W.~Graham, J.~Peraire, K.~Tang, Optimal control of vortex shedding using
  low-order models. part {I} -- open-loop model development, International
  Journal for Numerical Methods in Engineering 44~(7) (1999) 945--972.
\newblock \href
  {http://dx.doi.org/10.1002/(SICI)1097-0207(19990310)44:7<945::AID-NME537>3.0.CO;2-F}
  {\path{doi:10.1002/(SICI)1097-0207(19990310)44:7<945::AID-NME537>3.0.CO;2-F}}.

\bibitem{lions1973non}
J.~Lions, E.~Magenes, Non-homogeneous boundary value problems and applications.

\bibitem{Sirisup}
S.~Sirisup, G.~Karniadakis, Stability and accuracy of periodic flow solutions
  obtained by a {POD}-penalty method, Physica D: Nonlinear Phenomena 202~(3-4)
  (2005) 218--237.
\newblock \href {http://dx.doi.org/10.1016/j.physd.2005.02.006}
  {\path{doi:10.1016/j.physd.2005.02.006}}.

\bibitem{epshteyn2007estimation}
Y.~Epshteyn, B.~Rivi{\`e}re, {Estimation of penalty parameters for symmetric
  interior penalty Galerkin methods}, Journal of Computational and Applied
  Mathematics 206~(2) (2007) 843--872.
\newblock \href {http://dx.doi.org/10.1016/j.cam.2006.08.029}
  {\path{doi:10.1016/j.cam.2006.08.029}}.

\bibitem{kalashnikova2012efficient}
I.~Kalashnikova, M.~Barone, {Efficient non-linear proper orthogonal
  decomposition/Galerkin reduced order models with stable penalty enforcement
  of boundary conditions}, International Journal for Numerical Methods in
  Engineering 90~(11) (2012) 1337--1362.
\newblock \href {http://dx.doi.org/10.1002/nme.3366}
  {\path{doi:10.1002/nme.3366}}.

\bibitem{ferziger2002computational}
J.~H. Ferziger, M.~Peri{\'c}, Computational methods for fluid dynamics, Vol.~3,
  Springer, 2002.
\newblock \href {http://dx.doi.org/10.1063/1.881751}
  {\path{doi:10.1063/1.881751}}.

\bibitem{ITHACA}
G.~Stabile, G.~Rozza, {ITHACA}-{FV} - {I}n real {T}ime {H}ighly {A}dvanced
  {C}omputational {A}pplications for {F}inite {V}olumes,
  www.mathlab.sissa.it/ithaca-fv, accessed 2020-01-29.

\bibitem{niemann2016buoyancy}
M.~Niemann, J.~Fr{\"o}hlich, Buoyancy-affected backward-facing step flow with
  heat transfer at low prandtl number, International Journal of Heat and Mass
  Transfer 101 (2016) 1237--1250.
\newblock \href {http://dx.doi.org/10.1016/j.ijheatmasstransfer.2016.05.137}
  {\path{doi:10.1016/j.ijheatmasstransfer.2016.05.137}}.

\bibitem{Hess_CMAME_locROM}
M.~W. Hess, A.~Alla, A.~Quaini, G.Rozza, M.~Gunzburger, { A Localized
  Reduced-Order Modeling Approach for PDEs with Bifurcating Solutions},
  Computer Methods in Applied Mechanics and Engineering 351 (2019) 379 -- 403.
\newblock \href {http://dx.doi.org/10.1016/j.cma.2019.03.050}
  {\path{doi:10.1016/j.cma.2019.03.050}}.

\bibitem{star2019extension}
S.~K. Star, G.~Stabile, F.~Belloni, G.~Rozza, J.~Degroote, {Extension and
  comparison of techniques to enforce boundary conditions in {F}inite {V}olume
  {POD--G}alerkin reduced order models for fluid dynamic problems}\href
  {http://arxiv.org/abs/1912.00825} {\path{arXiv:1912.00825}}.

\bibitem{baiges2014reduced}
J.~Baiges, R.~Codina, S.~R. Idelsohn, {Reduced-order modelling strategies for
  the finite element approximation of the incompressible Navier-Stokes
  equations}, in: Numerical Simulations of Coupled Problems in Engineering,
  Springer, 2014, pp. 189--216.
\newblock \href {http://dx.doi.org/10.1007/978-3-319-06136-8\_9}
  {\path{doi:10.1007/978-3-319-06136-8\_9}}.

\bibitem{Akhtar}
I.~Akhtar, A.~H. Nayfeh, C.~J. Ribbens, On the stability and extension of
  reduced-order {G}alerkin models in incompressible flows, Theoretical and
  Computational Fluid Dynamics 23~(3) (2009) 213--237.
\newblock \href {http://dx.doi.org/10.1007/s00162-009-0112-y}
  {\path{doi:10.1007/s00162-009-0112-y}}.

\bibitem{bergmann2009enablers}
M.~Bergmann, C.-H. Bruneau, A.~Iollo, Enablers for robust {POD} models, Journal
  of Computational Physics 228~(2) (2009) 516--538.
\newblock \href {http://dx.doi.org/10.1016/j.jcp.2008.09.024}
  {\path{doi:10.1016/j.jcp.2008.09.024}}.

\bibitem{fick2017reduced}
L.~Fick, Y.~Maday, A.~T. Patera, T.~Taddei, A reduced basis technique for
  long-time unsteady turbulent flows, Book of Abstracts ENUMATH 2017 (2017) 70.

\bibitem{kunisch2010optimal}
K.~Kunisch, S.~Volkwein, {Optimal snapshot location for computing POD basis
  functions}, ESAIM: Mathematical Modelling and Numerical Analysis 44~(3)
  (2010) 509--529.
\newblock \href {http://dx.doi.org/10.1051/m2an/2010011}
  {\path{doi:10.1051/m2an/2010011}}.

\bibitem{drikakis2009large}
D.~Drikakis, M.~Hahn, A.~Mosedale, B.~Thornber, Large eddy simulation using
  high-resolution and high-order methods, Philosophical Transactions of the
  Royal Society A: Mathematical, Physical and Engineering Sciences 367~(1899)
  (2009) 2985--2997.
\newblock \href {http://dx.doi.org/10.1098/rsta.2008.0312}
  {\path{doi:10.1098/rsta.2008.0312}}.

\bibitem{grotzbach1999direct}
G.~Gr{\"o}tzbach, M.~W{\"o}rner, Direct numerical and large eddy simulations in
  nuclear applications, International Journal of Heat and Fluid Flow 20~(3)
  (1999) 222--240.
\newblock \href {http://dx.doi.org/10.1016/S0142-727X(99)00012-0}
  {\path{doi:10.1016/S0142-727X(99)00012-0}}.

\bibitem{simoneau2010applications}
J.-p. Simoneau, J.~Champigny, O.~Gelineau, Applications of large eddy
  simulations in nuclear field, Nuclear Engineering and Design 240~(2) (2010)
  429--439.
\newblock \href {http://dx.doi.org/10.1016/j.nucengdes.2008.08.018}
  {\path{doi:10.1016/j.nucengdes.2008.08.018}}.

\bibitem{argyropoulos2015recent}
C.~Argyropoulos, N.~Markatos, Recent advances on the numerical modelling of
  turbulent flows, Applied Mathematical Modelling 39~(2) (2015) 693--732.
\newblock \href {http://dx.doi.org/10.1016/j.apm.2014.07.001}
  {\path{doi:10.1016/j.apm.2014.07.001}}.

\bibitem{wang2012proper}
Z.~Wang, I.~Akhtar, J.~Borggaard, T.~Iliescu, Proper orthogonal decomposition
  closure models for turbulent flows: a numerical comparison, Computer Methods
  in Applied Mechanics and Engineering 237 (2012) 10--26.
\newblock \href {http://dx.doi.org/10.1016/j.cma.2012.04.015}
  {\path{doi:10.1016/j.cma.2012.04.015}}.

\bibitem{xie2018data}
X.~Xie, M.~Mohebujjaman, L.~G. Rebholz, T.~Iliescu, Data-driven filtered
  reduced order modeling of fluid flows, SIAM Journal on Scientific Computing
  40~(3) (2018) B834--B857.
\newblock \href {http://dx.doi.org/10.1137/17M1145136}
  {\path{doi:10.1137/17M1145136}}.

\bibitem{osth2014need}
J.~{\"O}sth, B.~R. Noack, S.~Krajnovi{\'c}, D.~Barros, J.~Bor{\'e}e, {On the
  need for a nonlinear subscale turbulence term in POD models as exemplified
  for a high-Reynolds-number flow over an Ahmed body}, Journal of Fluid
  Mechanics 747 (2014) 518--544.
\newblock \href {http://dx.doi.org/10.1017/jfm.2014.168}
  {\path{doi:10.1017/jfm.2014.168}}.

\bibitem{markatos1978transient}
N.~C. Markatos, Transient flow and heat transfer of liquid sodium coolant in
  the outlet plenum of a fast nuclear reactor, International Journal of Heat
  and Mass Transfer 21~(12) (1978) 1565--1579.
\newblock \href {http://dx.doi.org/10.1016/0017-9310(78)90012-1}
  {\path{doi:10.1016/0017-9310(78)90012-1}}.

\bibitem{peherstorfer2015dynamic}
B.~Peherstorfer, K.~Willcox, Dynamic data-driven reduced-order models, Computer
  Methods in Applied Mechanics and Engineering 291 (2015) 21--41.
\newblock \href {http://dx.doi.org/10.1016/j.cma.2015.03.018}
  {\path{doi:10.1016/j.cma.2015.03.018}}.

\bibitem{frank2020machine}
M.~Frank, D.~Drikakis, V.~Charissis, Machine-learning methods for computational
  science and engineering, Computation 8~(1) (2020) 15.
\newblock \href {http://dx.doi.org/10.3390/computation8010015}
  {\path{doi:10.3390/computation8010015}}.

\end{thebibliography}

\end{document}